\documentclass{article}
\usepackage{graphicx}
\usepackage{amsmath}
\usepackage{amsopn}
\usepackage{amsfonts}
\usepackage{amssymb}
\usepackage[square]{natbib}
\usepackage{url}
\usepackage{subfigure}
\usepackage{booktabs}
\usepackage{rotating}
\usepackage[hang,footnotesize,bf]{caption}
\usepackage{setspace}
\usepackage{multirow}
\usepackage[]{algorithm}
\usepackage{algorithmic}
\usepackage{epstopdf}

\usepackage{colortbl}
\arrayrulecolor{black}
\doublerulesepcolor{black}

\newcommand*{\IsDraft}{false}
\newcommand*{\fullwidth}{16.5cm}
\newcommand*{\threequarterwidth}{12.5cm}
\newcommand*{\mediumwidth}{10cm}
\newcommand*{\halfwidth}{8cm}

\newcommand{\be}{\begin{equation}}
\newcommand{\ee}{\end{equation}}
\newcommand{\ba}{\begin{eqnarray}}
\newcommand{\ea}{\end{eqnarray}}

\begin{document}
\title{High Resolution Long- and Short-Term Earthquake Forecasts for California}

\author{
Maximilian J.\ Werner\textsuperscript{1}\textsuperscript{*},
Agn\`{e}s Helmstetter\textsuperscript{2},
David D. Jackson\textsuperscript{3},
 and Yan Y. Kagan\textsuperscript{3}}
\date{\today}
\maketitle
\textsuperscript{1} Swiss Seismological Service, Institute of Geophysics, ETH Zurich, Switzerland.

\textsuperscript{2} Laboratoire de G\'eophysique Interne et Tectonophysique, Universit\'e Joseph Fourier and Centre National de la Recherche Scientifique, Grenoble, France.

\textsuperscript{3} Department of Earth and Space Sciences, University of California, Los Angeles, USA. 

\textsuperscript{*} Corresponding author:  \\
\hspace* {4cm} Maximilian J. Werner \\
\hspace* {4cm} Swiss Seismological Service \\
\hspace* {4cm} Institute of Geophysics\\
\hspace* {4cm} Sonneggstr. 5\\
\hspace* {4cm} 8092 Zurich, Switzerland\\
\hspace* {4cm} {\bf mwerner@sed.ethz.ch}

\doublespacing

\abstract

We present two models for estimating the probabilities of future earthquakes in California, to be tested in the Collaboratory for the Study of Earthquake Predictability (CSEP). The first, time-independent model, modified from \citet{Helmstetter-et-al2007}, provides five-year forecasts for magnitudes $m \geq 4.95$. We show that large quakes occur on average near the locations of small $m \geq 2$ events, so that a high-resolution estimate of the spatial distribution of future large quakes is obtained from the locations of the numerous small events. We employ an adaptive spatial kernel of optimized bandwidth and assume a universal, tapered Gutenberg-Richter distribution. In retrospective tests, we show that no Poisson forecast could capture the observed variability. We therefore also test forecasts using a negative binomial distribution for the number of events. We modify existing likelihood-based tests to better evaluate the spatial forecast. Our time-dependent model, an Epidemic Type Aftershock Sequence (ETAS) model modified from \citet{Helmstetter-et-al2006}, provides next-day forecasts for $m \geq 3.95$. The forecasted rate is the sum of a background rate, proportional to our time-independent model, and of the triggered events due to all prior earthquakes. Each earthquake triggers events with a rate that increases exponentially with its magnitude and decays in time according to Omori's law. An isotropic kernel models the spatial density of aftershocks for small ($\leq 5.5$) events. For larger quakes, we smooth early aftershocks to forecast later events. We estimate parameters by optimizing retrospective forecasts. Our short-term model realizes a gain of about 6.0 over the time-independent model.

\section{Introduction}

A wide range of ideas and hypotheses exist about how, when and where earthquakes occur and about how big they will be. 
Given seismicity's strong stochasticity, along with data quality and quantity issues, the most promising path towards weeding out the good ideas from the bad ones is rigorous, comparative and transparent evaluation of prospective, synoptic and probabilistic earthquake forecasts. Such forecasts make scientific hypotheses of earthquake occurrence testable, transparent and refutable. A major step along this path was taken by the Working Group on Regional Earthquake Likelihood Models (RELM), which invited long-term (5-year) forecasts for California in a specific format to facilitate comparative testing \citep{Field2007, Schorlemmer-et-al2007, SchorlemmerGerstenberger2007, Schorlemmer-et-al2009}. Building on RELM's success, the Collaboratory for the Study of Earthquake Predictability (CSEP, \url{www.cseptesting.org}) inherited and expanded RELM's mission to regionally and globally test prospective forecasts \citep{Jordan2006, Schorlemmer-et-al2009, Zechar-et-al2009}. 

The development of better models to be tested in CSEP remains the highest priority. We present two models to estimate the probabilities of future earthquakes in California. The first model provides time-independent, long-term (5-year) estimates of the probabilities of future earthquakes of magnitudes $m \ge 4.95$ in California, in a format suitable for testing within CSEP. The second model estimates short-term (one-day) probabilities of future $m \ge 3.95$ earthquakes. Both models are extended and/or modified versions of those developed by \citet{Helmstetter-et-al2007} and \citet{Helmstetter-et-al2006}. We made improvements to the long-term model and updated using the latest available data. The short-term model was extended from southern California to all of California. We made some further modifications (described in the text) and re-estimated its parameters using all available data up to 1 April 2009. 

Both models are based on simple, yet hotly contested hypotheses. Both models solely require past seismicity as data input. Whether and at which point other data such as geological or geodetic data could improve the forecasts remains an interesting open question. The long-term model assumes that future earthquakes will occur with higher probability in areas where past earthquakes have occurred. To turn this hypothesis into a testable forecast, we smoothed the locations of past seismicity using an adaptive kernel. The long-term model assumes that sequences of triggered events are mostly temporary perturbations to the long-term rate; we therefore attempted to remove them using a relatively arbitrary method. Moreover, the magnitude of each earthquake is independently distributed according to a tapered Gutenberg-Richter distribution with corner magnitude $m_c=8.0$ relevant for California \citep{BirdKagan2004}, independent of geological setting, past earthquakes or any other characteristic.  Because of the assumed magnitude independence, the model also uses small $m>2$ events to forecast large $m>4.95$ events. The small quakes indicate regions of active seismicity, and retrospective tests support the claim that including these small events improves forecasts of larger ones. 

The time-dependent, short-term model makes similar assumptions. First, the same assumption holds regarding the distribution of magnitudes. Second, every earthquake can trigger future earthquakes with a rate that increases exponentially with the magnitude of the triggering event. The subjective, retrospective distinction between fore-, main- and aftershocks is thereby eliminated: every earthquake can trigger other events, and those triggered events may be larger than the triggering shock. The time-dependence enters in the form of the well-known Omori-Utsu law. However, in our model the Omori-Utsu law applies to all earthquakes, not just to large earthquakes that trigger smaller ones. In addition to the contribution to the seismicity from past quakes, the model assumes that there exists a background rate, which is modeled as a spatially heterogeneous Poisson process. The background's normalized spatial distribution is taken directly from the first, time-independent model, but its rate is estimated. The model is a particular implementation of the epidemic-type aftershock sequence (ETAS) model \citep{Ogata1988}. The model may be interpreted as a simple but powerful null hypothesis of earthquake clustering and triggering, based on empirical distributions of seismicity.  


This article (and its five electronic supplements) describes the two models, their calibration on earthquake catalogs, results from retrospective forecasts, and what we learned about seismicity in the process. With these forecasts, we respond to CSEP's call for testable forecasts in a specific format: The expected number of quakes in individual magnitude bins of width $0.1$ from $m=4$ to $m=9$ in each spatial cell of $0.1$ by $0.1$ degrees latitude/longitude in a predefined testing region that includes California and extends about one degree beyond its borders \citep{SchorlemmerGerstenberger2007}.  

The article is structured as follows. We describe the data in section \ref{sec:Data} (removing explosions is discussed in Supplement 1). Section \ref{sec:Spont} explains our method for estimating the spatial distribution of spontaneous seismicity, which we used both for the long-term forecast and as the background in the short-term forecasts. We separated the data into two sets: the first set (the learning or training catalog) is used to forecast the second set (the target catalog). We estimated smoothing parameters by optimizing these forecasts. In Supplement 2, we optimize parameters for different target catalog magnitude thresholds. In section \ref{sec:TI}, we calculate the expected number of events and apply the Gutenberg-Richter distribution to generate the time-independent five-year forecast. In  Supplement 3, we discuss a forecast for the so-called CSEP mainshock-only forecast group. To assess the time-dependent forecasts, we rescaled the five-year forecast to a one-day time-independent forecast (section \ref{sec:fc2}). Section \ref{sec:ETAS} defines the ETAS model, describes the parameter estimation and discusses the model's goodness of fit to the data. Supplement 4 treats the performance of the parameter optimization algorithm. Supplement 5 provides parameter estimates for target magnitudes $m \geq 2$ rather than the threshold $m\geq3.95$ of the CSEP one-day forecast group. Section \ref{sec:TD} compares the ETAS forecast with the time-independent model during the Baja swarm of February 2009. Both models will be installed at the SCEC testing center of CSEP.


\section{Data: The ANSS Earthquake Catalog}
\label{sec:Data}
We used the ANSS catalog in the period from 1 January 1981 until 1 April 2009 with magnitude $ m \geq 2$ in the spatial region defined by the RELM/CSEP collection region, as defined by the polygon in Table 2 by \citet{SchorlemmerGerstenberger2007}. The catalog listed $167,548$ events. Underground nuclear explosions at the Nevada Test Site contaminate the earthquake catalog. Supplement 1 documents how we identified and removed 21 explosions from the catalog, 9 of which were large $m \geq 5$ events. 

Earthquake catalogs change over time, because of reprocessing, deletion or addition of past events. As a result, forecasts based on supposedly identical data differ from each other, sometimes significantly. We encountered this problem when attempting to reproduce the results of \citet{Helmstetter-et-al2007}. The sole way to guarantee full reproducibility is thus to store the data set. The interested reader may obtain the data (and computer programs) from \url{mercalli.ethz.ch/~mwerner/WORK/CaliforniaForecasts/CODE/}.


\section{Spatial Distribution of Spontaneous Seismicity}
\label{sec:Spont}

\subsection{Declustering Seismicity}

\begin{figure}
\centering
\includegraphics[draft=\IsDraft,width=\threequarterwidth,keepaspectratio=true,clip]{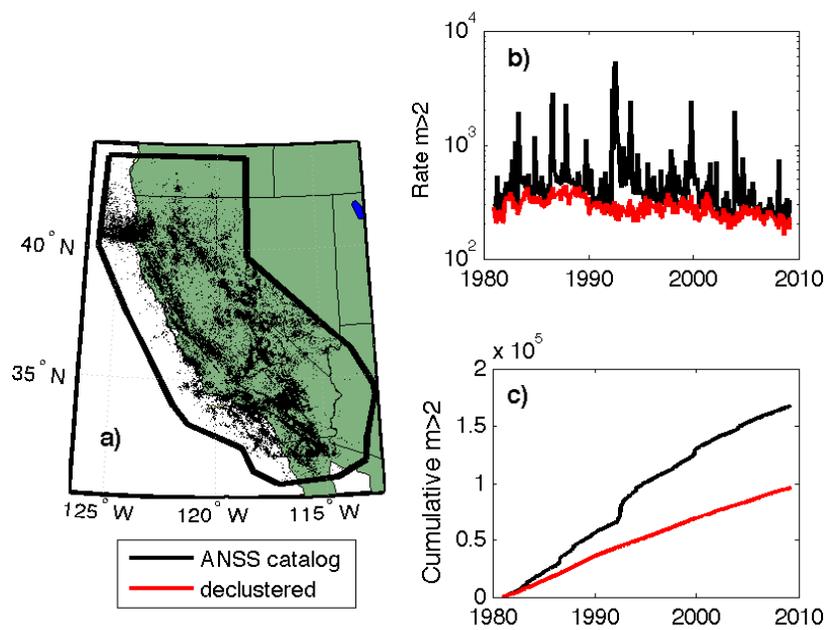}
\caption{\label{Fig:DeclusterSpace} a): Original ANSS catalog $m \geq 2$ from 1 January 1981 until 1 April 2009 in the CSEP collection polygon around California. b): Original (black) and declustered (red) seismicity per month. c): Original (black) and declustered (red) cumulative seismicity. To decluster, we used a method modified from \citet{Reasenberg1985}.
}
\end{figure}

To estimate the spatial distribution of independent events, we used a modified version of the Reasenberg declustering algorithm \citep{Reasenberg1985,Helmstetter-et-al2007}. We set the input parameters to $r_{fact} = 8$ (dimensionless), $x_{meff}=2$ (units of magnitude), $x_k=0.5$ (dimensionless), $p_1=0.95$ (dimensionless probability), $\tau_{min}=1$ day and $\tau_{max}= 5$ days. We used as the interaction distance the scaling $r=0.01 \times 10^{0.5M}$ km, as suggested by \citet{WellsCoppersmith1994}, instead of $r=0.011 \times 10^{0.4M}$ km and $r < 30$ km in Reasenberg's algorithm. We further set the location errors to $1$ km horizontal and $2$ km vertical. Figure \ref{Fig:DeclusterSpace} compares the original with the declustered catalog. The declustering algorithm found that $57 \%$ of the events were spontaneous. We chose this method to estimate spontaneous seismicity over more sophisticated algorithms based on stochastic process theory\citep{Kagan1991a, Zhuang-et-al2002, Zhuang-et-al2004, MarsanLengline2008} because of its simplicity. The declustering algorithm was not optimized for forecasting  and the particular procedure may have significant effects on the forecasts. In the future, we'd like to test and optimize the declustering algorithm, too. 

\subsection{Adaptive Kernel Smoothing of Declustered Seismicity}
\label{sec:AK}

We estimated the density of spontaneous seismicity in each $0.1$ by $0.1$ degree cell by smoothing the location of each earthquake in the declustered catalog with an isotropic adaptive kernel $K_{d_i} (\vec{r})$. We tested two choices for $K_{d} (\vec{r})$, either a power-law
\be
K_{d} \left(\vec{r}\right) = \frac{C(d)}{\left(|\vec{r}|^2+d^2\right)^{1.5}}
\label{eq:Kpl}
\ee
or a Gaussian
\be
K_{d} \left(\vec{r}\right) = C'(d) \exp{\left[ - \frac{|\vec{r}|^2}{2 d^2} \right]}
\label{eq:Kgauss}
\ee
where $d$ is the adaptive smoothing distance, and $C(d)$ and $C'(d)$ are normalizing factors, so that the integral of $K_{d} \left(\vec{r}\right)$ over an infinite area equals 1. We measured the smoothing distance $d_i$ associated with an earthquake $i$ as the horizontal (epicentral) distance between event $i$ and its $n_v$th closest neighbor. The number of neighbors, $n_v$, is an adjustable parameter, estimated by optimizing the spatial forecasts (see section \ref{sec:smoothopt}). We imposed $d_i \geq 0.5$ km to account for location uncertainty. The kernel bandwidth $d_i$ thus decreases in regions of dense seismicity, so that we have better resolution (smaller $d_i$) where the density is higher. 

The density $\mu(\vec{r})$ at any point $\vec{r}$ is then estimated from the $N$ events by
\be
\mu(\vec{r}) = \sum_{i=1}^N K_{d_i} \left(\vec{r}-\vec{r_i} \right)
\label{eq:mu}
\ee
However, our forecasts are given as an average number of events in each $0.1^\circ$ by $0.1^\circ$ cell. We therefore integrated equation (\ref{eq:mu}) over each cell to obtain the seismicity rate in this cell. 

\subsection{Correcting for Spatial Magnitude Incompleteness}
\label{sec:mcSpace}

We used events with magnitudes $m \geq 2$ to estimate the spatial distribution of seismicity. Unfortunately, the catalog is not complete everywhere at this magnitude level. To correct for catalog incompleteness, we estimated the completeness magnitude $m_0$ in each cell from the magnitude distribution. 

\begin{figure}
\centering
\includegraphics[draft=\IsDraft,width=\threequarterwidth,keepaspectratio=true,clip]{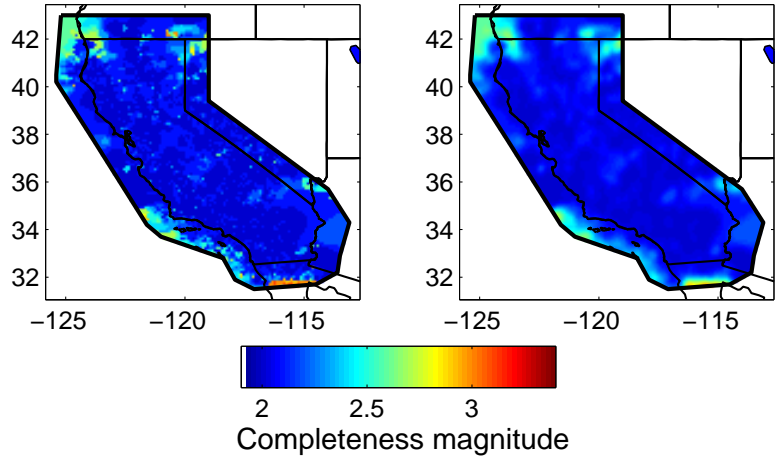}
\caption{\label{fig:mc} Left: Completeness magnitude estimated from the maxima of the magnitude distribution in each cell. Right: Smoothed completeness magnitude.
}
\end{figure}

We can estimate the magnitude distribution $P_m(\vec{r},m)$ at point $\vec{r}$ using the same kernel method as described above to estimate the density $\mu(\vec{r})$. We smoothed both the locations and the magnitudes of all (declustered) earthquakes using
\be
P_m\left(\vec{r},m\right) = \sum_{i=1}^N K_{d_i} \left( \vec{r} - \vec{r_i} \right) G_h \left( m-m_i \right)
\label{eq:pmr}
\ee
where $G_h(m)$ is a Gaussian function of mean $m$ and width $h$. The kernel width $h$ was fixed to $0.15$ magnitude units. We then integrated $P_m(\vec{r},m)$ over each cell to get the magnitude distribution in these cells. We estimated the completeness magnitude $m_0$ in each cell as the magnitude at which the smoothed magnitude distribution is at a maximum. 
Using this method, we obtain large small-scale fluctuations of $m_0$, which are probably unphysical. The completeness magnitude should be relatively smooth at scales smaller than the typical distance between seismic stations. Therefore, we smoothed the completeness magnitude $m_0$ using a Gaussian filter with a standard deviation of $0.15^\circ$. The result is shown in Figure \ref{fig:mc}, where we used the power-law kernel (\ref{eq:Kpl}) with a smoothing distance $d_i$ in equation (\ref{eq:pmr}) to the 6th nearest neighbor (i.e. $n_v=6$). Most of the region has $m_0 \approx 2$ (our method does not estimate completeness thresholds smaller than $m=2$). The completeness magnitude is much larger, close to $m_0=3.5$, close to the boundaries of the collection region, especially in the Mendocino area and in Mexico. Comparing our Figure \ref{fig:mc} to Figure 2 of \citet{Helmstetter-et-al2007}, we found significant differences in the estimated completeness magnitude due to two reasons. Firstly, we found and corrected a bug in the smoothing algorithm of \citet{Helmstetter-et-al2007}, which artificially increased the completeness magnitude near the boundaries. Secondly, we used more earthquake data, up to 1 April 2009. 

Our method failed when few earthquakes exist from which to estimate the completeness magnitude. We introduced two simple ad hoc rules for such situations. First, if the magnitude threshold of the learning catalog is $m_{min} \geq 3.0$, we set the completeness threshold to $m_0=3.0$ to eliminate clearly unphysical fluctuations. Thus, we only correct for a spatially varying completeness threshold for learning catalogs that include events $m_{min}<3$, when completeness effects are severe. Second, if the estimated $m_0 \geq 3.5$, we set $m_0=3.5$ to make sure the completeness level remains reasonable. There are other, more sophisticated methods to estimate the completeness magnitude (see, e.g., \citep{SchorlemmerWoessner2008} and references therein), which we hope to use in the future. 

Since we are interested in estimating the rate of earthquakes $m \geq m_{min}$, we corrected the observed rate for missing events by applying a Gutenberg-Richter scaling \citep{GutenbergRichter1944} with $b=1$ (see section \ref{sec:MD} for the b-value estimation)
\be
\mu'(\vec{r}) = \mu(\vec{r}) \cdot 10^{m_0(\vec{r}) - m_{min}}
\label{eq:mum}
\ee

\subsection{Optimizing the Spatial Smoothing}
\label{sec:smoothopt}

We estimated the parameter $n_v$, the number of neighbors used to compute the smoothing distance $d_i$ in equation (\ref{eq:mu}), by maximizing the likelihood of the model. We built the model $\mu'(i_x,i_y)$ in each cell $(i_x,i_y)$ on a training period of the catalog and tested it (evaluated the likelihood) on a separate testing period of the catalog. Our model assumes independence of magnitudes and locations. We therefore solely tested the spatial distribution of a model forecast, neglecting magnitudes. Moreover, we assumed that spontaneous background seismicity is time-independent, so that the expected number of events is simply a pre-factor to a normalized spatial density. Therefore, we optimized solely the normalized spatial density estimate in each cell $(i_x,i_y)$ using
\be
\mu^*(i_x,i_y) = \frac{\mu'(i_x,i_y) N_t}{\sum_{i_x} \sum_{i_y}\mu'(i_x,i_y)}
\label{eq:munorm}
\ee
where $N_t$ is the number of observed target events. The expected number of events for the model $\mu^*$ thus equals the observed number $N_t$ to optimize solely the spatial forecast. 

We assumed that the observed earthquakes in each cell are Poisson random variables, independent of each other in space and time. We revisit this assumption in sections \ref{sec:retro} and \ref{sec:NBD}. 
The log-likelihood of the model is thus given by
\be
LL = \sum_{i_x} \sum_{i_y} \log p \left[ \mu^* (i_x,i_y), n \right]
\label{eq:L}
\ee
where $n$ is the number of events that occurred in cell $(i_x,i_y)$, and the probability $p$ of observing $n$ events in cell $(i_x,i_y)$ given a forecast of $\mu^* (i_x,i_y)$ in that cell is given by the Poisson distribution
\be
p \left[ \mu^* (i_x,i_y), n \right] = \left[ \mu^*(i_x,i_y)  \right]^n \frac{\exp \left[ -\mu^*(i_x,i_y) \right]}{n!} \ .
\label{eq:Poi}
\ee
We built an extensive set of background models $\mu^*$ by varying (i) the spatial smoothing kernel, either Gaussian or power-law, (ii) the input/training catalog, and (iii) the target catalog. The training and the target catalog were chosen so that they do not overlap (except in certain cases discussed below). We generated forecasts based on the training catalog and evaluated them on the separate target catalog to test our forecasts out-of-sample and to avoid over-fitting. We evaluated the performance of each model by calculating its probability gain per earthquake relative to a model with a uniform spatial density:
\be
G = \exp \left( \frac{LL - LL_{unif}}{N_t} \right)
\label{eq:G}
\ee
where $LL_{unif}$ is the log-likelihood of the uniform model and $N_t$ is the number of observed events in the duration $t$ of the target catalog.

\subsection{Results of the Spatial Smoothing Optimization}
\label{sec:resultsopt}

We smoothed the declustered catalog according to the procedure described above and as previously performed by \citep{Helmstetter-et-al2007}. Apart from applying more (updated) data than \citet{Helmstetter-et-al2007} and using a modified procedure for estimating the completeness magnitude, we also corrected the code for artificial boundary effects and rounding errors which led to some events being assigned to the wrong cells. In the following sections, we highlight different aspects of the results of the optimization, which are summarized in Table \ref{tab:SmoOpt}. 

Table \ref{tab:SmoOpt} shows the probability gain of the smoothed seismicity forecasts as a function of the magnitude threshold $m_{min}$ of the learning catalog. The probability gain increases slightly from $m_{min}=2$ ($G=5.13$) until $m_{min}=3$ ($G=6.73$) before decreasing rapidly beyond $m_{min}=4$. Therefore, small $m \geq 2$ earthquakes help predict the locations of future large $m\geq 5$ shocks. 

While small earthquakes continue to help make better forecasts, the precise values of the gain fluctuate for different target periods. Smaller gains indicate that some target events are surprises, occurring in locations of little previous seismicity. Such events may be better forecast by a longer catalog. For the 1989 to 1993 target period, the poor likelihood score was due to three earthquakes that occurred near $(42.3N, 122.0E)$. There, the uniform forecast ($\mu_{unif}=1.3 \cdot 10^{-4}$ expected quakes per 5 years per cell) outperformed the smoothed seismicity forecast ($\mu \approx 3 \cdot 10^{-6}$ expected quakes per 5 years per cell). Thus, large, surprising events may occur in regions that saw no activity in over a decade, not even small $m\geq2$ events. 

Interestingly, the gain does not monotonically decrease with shorter learning period, i.e. statistical fluctuations influence the results. Nevertheless, including small earthquakes to generate forecasts for large $m>5$ earthquakes improves estimates of earthquake potential over estimates based solely on large events. This suggests that large earthquakes occur on average in the same locations as small earthquakes.


We tested two spatial smoothing kernels: the Gaussian and the power-law kernel. Table \ref{tab:SmoOpt} shows that the Gaussian kernel performs marginally better than the power-law kernel, in contrast to the results obtained by \citet{Helmstetter-et-al2007}. However, different target periods can show reversed results. Given that the gain varies between different target periods, the differences between the Gaussian and the power-law kernel may not be significant. We preferred the power-law kernel because seismicity occurs on fractal networks and because the optimal smoothing parameter $n_v$ was more robust. 

In Table \ref{tab:SmoOpt}, we tested the influence of varying the magnitude threshold of the learning catalog. In Supplement 2, we show that the gain is roughly constant for models in which only the magnitude threshold of the target catalog is increased from $2$ to $5.5$, suggesting that large earthquakes on average occur in the same locations as small earthquakes. Our method works well for any target magnitudes. 


Our favorite spatial background for a new five-year forecast is given by model 21, which uses the entire available catalog. Compared to model 20, where $n_v=1$, model 21 had slightly smaller gain. However, the optimal smoothing parameter $n_v$ varies from $1$ (model 3) to $9$ (model 19) for different 5-year target periods. The average optimal $n_v$ over the 4 different target periods equals 3.75, which rounds to $4$. However, a slightly smoother model is preferable since we only used a short catalog since 1981, while maintaining a comparable probability gain. Model 21 with $n_v=6$ fulfilled these requirements. In the future, we'd like to optimize the smoothing parameter $n_v$ over different target periods. 


In Supplement 3, we repeated the spatial forecast optimization on a declustered target catalog, in order to generate a second long-term forecast targeting the mainshock-only forecast group of CSEP \citep{Schorlemmer-et-al2007, SchorlemmerGerstenberger2007}. 

We performed further tests to investigate the effects of different input and target catalogs. Increasing the collection region from the CSEP collection region to a much larger rectangle around California slightly increases the performance of the smoothed seismicity forecasts. 
A spatial forecast produced by smoothing the entire non-declustered, original training catalog performed significantly worse than the declustered smoothed seismicity. Without accounting for the temporal Omori-Utsu decay of clustered events, the long-term forecast is dominated by short-term clustering and too localized.

\section{Time-Independent 5-Year $m\geq 4.95$ Forecast Based on Smoothed Seismicity}
\label{sec:TI}

\subsection{Magnitude Distribution}
\label{sec:MD}

We assumed that the cumulative magnitude probability distribution obeys a tapered Gutenberg-Richter distribution with a uniform b-value and corner magnitude $m_c$ (Eq. 10 in \citet{Helmstetter-et-al2007})
\begin{equation}
P(M>m)=10^{-b(m-m_{min})} \exp\left[ 10^{1.5(m_{min}-m_c)}-10^{1.5(m-m_c)}\right]
\label{tGR}
\end{equation}
with a minimum magnitude $m_{min}=4.95$ (for the five-year CSEP forecast group) and a corner magnitude $m_c=8.0$ as suggested by \citet{BirdKagan2004} for continental transform fault boundaries. We estimated the b-value using maximum likelihood \citep{Aki1965} and found $b\approx1$ for $m\geq 2$. 

\begin{figure}
\centering
\includegraphics[draft=\IsDraft,width=\halfwidth,keepaspectratio=true,clip]{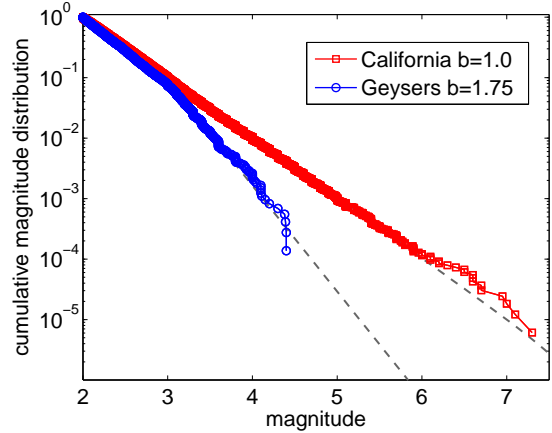}
\caption{\label{fig:geysers} Magnitude distribution of earthquakes $m\geq 2$ from 1 January 1981 to 1 April 2009 in the Geysers region (blue circles) and in the remainder of California (red squares). Near the geothermally active Geysers, the distribution differs from the usual Gutenberg-Richter distribution with b-value equal to one. 
}
\end{figure}

In the geothermally active Geysers region in Northern California, the distribution of the magnitudes deviates from the standard Gutenberg-Richter distribution with $b$ equal to one (Figure \ref{fig:geysers}): a break occurs around $m\simeq 3.3$. Below the break, the b-value is $b\approx1$, while above it is $b\approx1.75$. To forecast the expected number of earthquakes in each magnitude bin of size $\Delta m=0.1$, we used this modified magnitude distribution for the Geysers area (for $-122.9^\circ < $lon$ < -122.7^\circ$ and $ 38.7^\circ < $lat$ < 38.9^\circ$). Other geothermal regions in California seem to behave more regularly, indicating that geothermal activity is not the sole cause of the anomalous magnitude distribution at the Geysers \citep{Helmstetter-et-al2007}. 

\subsection{Expected Number of Events}
\label{sec:ExpNum}

To estimate the expected number of earthquakes, we counted the number of $m\geq 4.95$ events in the time period from 1 January 1981 until 1 April 2009 and divided by the time period of the catalog in years to obtain $N_{pred}=6.71$ earthquakes per year in the CSEP testing region. The expected number of events per year in each space-magnitude bin $\left(i_x,i_y,i_m \right)$ was then calculated from
\begin{equation}
E\left(i_x,i_y,i_m \right)= N_{pred} \ \mu(i_x,i_y) \ P(i_m)
\label{eq:expEv}
\end{equation}
where $\mu$ is the normalized spatial background density, and $P(i_m)$ is the integrated probability of an earthquake in magnitude bin $(i_m)$ defined according to equation (\ref{tGR}) and taking into account the different b-value for the Geysers. 

\begin{figure}
\centering
\includegraphics[draft=\IsDraft,width=\fullwidth,keepaspectratio=true,clip]{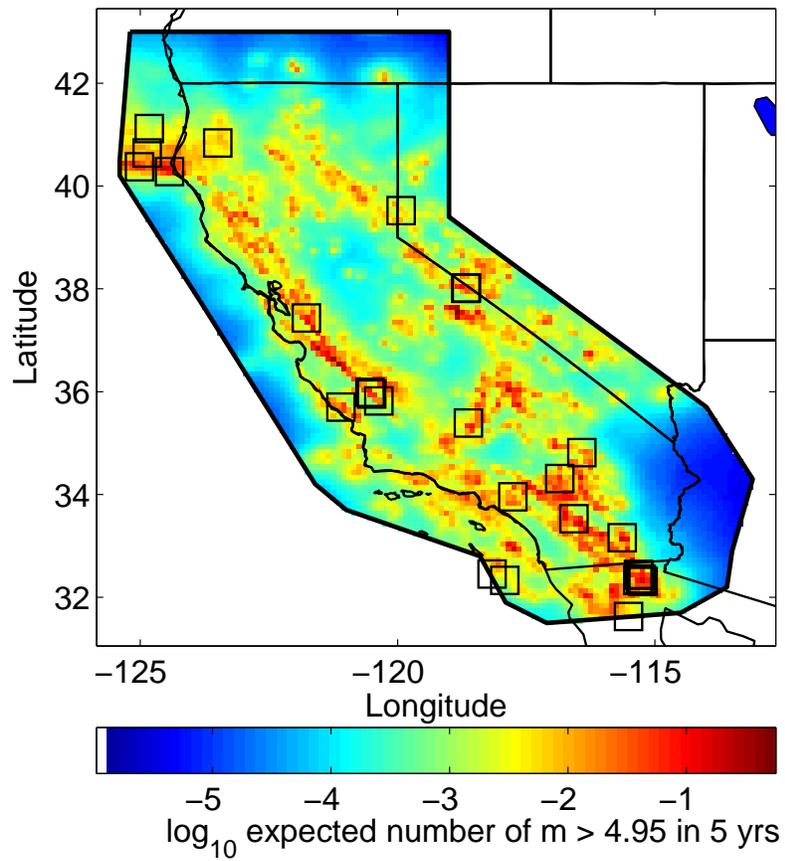}
\caption{\label{fig:new5yr} Expected number of earthquakes $m\geq 4.95$ per $0.1^o$ by $0.1^o$ cell per five years from 2010 to 2015, based on model 21 in Table \ref{tab:SmoOpt}. We used an optimized, adaptive power-law kernel to smooth the locations of declustered $m\geq2$ earthquakes from 1 January 1981 until 1 April 2009. Also shown as black squares are $m\geq 4.95$ quakes that occurred between 2003 and 2008. 
}
\end{figure}

Figure \ref{fig:new5yr} shows our new five-year forecast for $m\geq 4.95$ for 2010 to 2015 based on optimally and adaptively smoothing the locations of declustered, small $m\geq 2 $ earthquakes. On average, we expect $33.55$ earthquakes $m\geq 4.95$ over the next five years in the entire test region. The modified magnitude distribution near the Geysers helps us avoid an unrealistically high rate of large quakes there.

\subsection{Comparison of Our New Five-Year Forecast to \citet{Helmstetter-et-al2007}'s Forecast}
\label{sec:comp}

\begin{figure}
\centering
\includegraphics[draft=\IsDraft,width=\mediumwidth,keepaspectratio=true,clip]{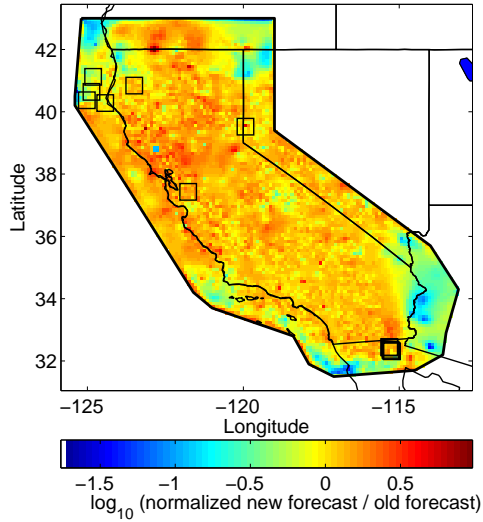}
\caption{\label{fig:comp}  Ratio of our new spatial five-year forecast for $m\geq 4.95$ over the spatial forecast by \citet{Helmstetter-et-al2007}, each normalized to $1$. Also shown by black squares are the 12 earthquakes $m\geq 4.95$ that occurred from 1 January 2006 until 1 July 2008, the half-time mark of the five-year prospective Regional Earthquake Likelihood Models (RELM) experiment, in which the model by \citet{Helmstetter-et-al2007} is leading (see \citep{Schorlemmer-et-al2009}). Five of the 12 events are located in Baja and overlap. 
}
\end{figure}

In generating our new five-year earthquake forecast, we used data up to 1 April 2009. \citet{Helmstetter-et-al2007} used data up to August 2005. We expect $33.55$ earthquakes $m\geq 4.95$ over the next five years, while \citet{Helmstetter-et-al2007} expected $35.40$ earthquakes $m\geq 4.95$ over the five-year period from January 2006 until December 2010. Another difference between our forecast and that of \citet{Helmstetter-et-al2007} is that we fixed two minor bugs in the smoothing algorithm and modified the procedure for estimating the completeness magnitude. To compare solely the spatial distribution of the forecasts, we normalized each forecast to an overall rate equal to $1$. Figure \ref{fig:comp} shows the ratio of the normalized new forecast over the normalized forecast by  \citet{Helmstetter-et-al2007}. 

Our new forecast is not very different from the old forecast in much of California (see the yellow regions in Figure \ref{fig:comp}). The differences (up to a factor of 10) in some localized regions are mostly due to recent earthquakes since August 2005 that increased the new forecast. However, there are also more unexpected differences. Firstly, the narrow stripes of increases or decreases along the boundaries were artifacts in the old forecast, which we fixed by correctly smoothing the completeness magnitude. Secondly, some regions in the South, South-East and in the North show that the old forecast expected about 30 times as many earthquakes as our new forecast. These differences are partially due to the corrected smoothing and partially due to the updated data set. 
The ANSS catalog may also have changed in the meantime, making forecast calculations somewhat irreproducible (see section \ref{sec:Data}). We stored the data at \url{mercalli.ethz.ch/~mwerner/WORK/CaliforniaForecasts/CODE/}.

Another region showing a factor $10$ difference is the Geysers area. \citet{Helmstetter-et-al2007} estimated $b \approx 1.99$ in this region, while we estimated $b\approx 1.75$. We also used a slightly different technique to extrapolate the observed $m>2$ rate to events $m>4.95$. 

The forecast by \citet{Helmstetter-et-al2007} was submitted to the five-year RELM experiment \citep{Field2007}. \citet{Schorlemmer-et-al2009} evaluated the submitted models after the first two and a half years of the five-year period. 12 earthquakes $m\geq4.95$ had been observed (see the black squares in Figure \ref{fig:comp}, five of which overlap in Baja California). The model by \citet{Helmstetter-et-al2007} outperformed all other models.  Our forecast, which was partially built on the same data set used to evaluate the forecast by \citet{Helmstetter-et-al2007}, would have outperformed the old forecast by a gain of $G=1.18$ per event.

How would our forecast compare if it were built on the same training data? We created a forecast based on the same data that \citet{Helmstetter-et-al2007} used, up to August 23, 2005. We compared both the goodness of fits and the performance during the first half of the RELM experiment. To compare goodness of fit, our model 22 in Table \ref{tab:SmoOpt} should be compared with model 21 in Table 1 of \citet{Helmstetter-et-al2007}: Our optimal smoothing parameter was $n_v=1$, while that of \citep{Helmstetter-et-al2007} was $n_v=6$; the numbers of earthquakes vary by several dozen (see section \ref{sec:Data}); the gain we found for $n_v=6$ is $G=6.85$, below the $G=7.08$ found by \citet{Helmstetter-et-al2007}, but for $n_v=1$, we find the same value of $G=7.08$. Thus, despite the minor bugs we identified and fixed, the goodness of fit is essentially the same. 

To compare performance in the RELM experiment, we needed to estimate the magnitude distribution and the number of expected events. We estimated $b\approx1$ for California. For the Geysers region, we found $b\approx1.88$, closer to the estimate $b\approx1.99$ by \citet{Helmstetter-et-al2007} than our previous one of $b\approx1.75$, for which we used all data up to April 2009. For the first half of the five-year RELM experiment period, the new forecast performed worse than the previous one ($G=0.83$ per earthquake). The differences are due to three events near the Mendocino Triple Junction. However, the differences are small and future earthquakes can change the probability gain significantly.

\subsection{Retrospective Consistency Tests of the 5-Year Forecast}
\label{sec:retro}

Thus far we optimized our model by maximizing its likelihood in retrospective forecasts, but we did not test whether our forecast is consistent with past data. To do that, we used the consistency tests of the ongoing RELM project \citep{Schorlemmer-et-al2007, Schorlemmer-et-al2009} (see also \citet{KaganJackson1994, Jackson1996}). We performed two different tests: one to test whether the expected number of events is consistent with the observed number of events (the N test), and another one to test whether the model's simulated likelihood values are consistent with the observed likelihood value (the L-test). 

Time-independent forecasts are usually specified as a Poisson rate, i.e. the probability distribution of the number of events under the model is given by a Poisson distribution with mean equal to the expected (mean) number of events ($N_{pred}=33.55$, in our case). Using this distribution, the N-test calculates the probability $\delta$ that, according to the model, the actually observed number or a smaller number is observed. If $\delta$ is very low (less than $0.025$) or very large (larger than $0.975$), we reject the model with $95\%$ confidence because, under the model, the observed number of events is highly unlikely. For the 5-year target period 2004-2008, there were $N_{obs}=25$ events, so that $\delta=0.08$ and the model could not be rejected.


To test whether the model is consistent with the observed likelihood value, we resorted to simulations. In each space-magnitude bin, we simulated the number of observed events according to the model forecast in that bin, again using the Poisson distribution. We then calculated the likelihood value of this particular simulation. We performed $10,000$ simulations, from which we obtained a distribution of likelihood values. The L test compares the observed log-likelihood value with these simulated log-likelihood values by calculating the probability $\gamma$ that the observed log-likelihood value or a smaller value was simulated. If the fraction $\gamma$ is extremely low (below $0.025$), then we reject the model, because, under the model, the observed log-likelihood score is highly unlikely. However, we do not reject the model if $\gamma$ is very large, since the model might be too smooth, so that the likelihood of the data might be much higher than expected under the model (see also \citet{Schorlemmer-et-al2007}). For the 5-year target period from 2004 to 2008 inclusive, the observed likelihood value was $LL_{obs}=-172.8$, which resulted in $\gamma=0.975$, i.e. the model was not rejected. 

\begin{figure}
\centering
\includegraphics[draft=\IsDraft,width=\halfwidth,keepaspectratio=true,clip]{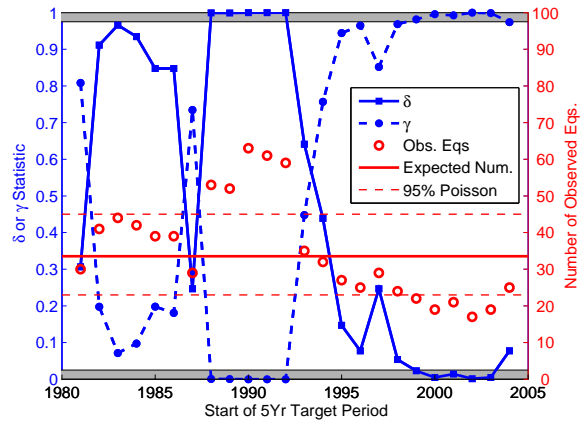}
\caption{\label{fig:relmRetro}  Retrospective consistency tests of our 5-year time-independent forecast on past, overlapping 5-year periods with starting year as abscissa. Left (blue) ordinate: The N-test statistic $\delta$ (blue squares), measuring the agreement between the observed and the predicted total number of events, and the L-test statistic $\gamma$ (blue circles), measuring the agreement between the observed and the expected likelihood score. Right (red) ordinate: Number of observed events in each target period (red circles), the expected number of events (red solid line) and $95\%$ Poisson confidence bounds (red dashed lines). Grey bars denote N-test rejection regions; the lower grey bar marks the L-test rejection region.
}
\end{figure}

We repeated the N and L tests of our forecast using a total of 24 overlapping 5-year target periods, the first of which started in 1981, the second in 1982, etc., until the last target period from 2004-2008. Figure \ref{fig:relmRetro} shows the $\delta$ and $\gamma$ values as a function of starting year of each of these 24 target periods. We also show for reference the number of observed events (red circles) in each target period, and our expected number of events $N_{pred}=33.55$ (solid red line) and the $95\%$ confidence bounds of the Poisson distribution (red dashed lines).

The model was rejected by the N test in several target periods: in those that contain the 1992 $M_W7.3$ Landers earthquake (too many observed events) and in a few that happen during the relatively quiet period 1999-2003 (too few observed events, despite the 1999 $M_W7.1$ Hector Mine sequence). The $\gamma$ statistic also rejected the model in the target periods that contain the 1992 $M_W7.3$ Landers earthquake. Since the L test is defined to be one-sided, we could not reject the model for $\gamma \ge 0.975$ during the quiet phase in the first years of the new millennium (but the N test did reject the model). The anti-correlations between the two statistics were very strong and indicate that, in this case, the L test is dominated by the number of observed events. To make the test more sensitive to the space and magnitude forecast of the model, we propose to modify the L-test in two ways: To normalize by the number of events, and to test only the spatial component of the forecast (see section \ref{sec:NBD}). \citet{Zechar-et-al2009b} also proposed the latter spatial (S) test and we follow their notation.

Figure \ref{fig:relmRetro} also shows that the assumption of Poissonian variability in the number of observed events is wrong: The variance of the number of observed events per 5 year period is larger than expected from the Poissonian model. The model is rejected ten times in 24 overlapping periods. Moreover, since the number of observed events sometimes exceeded the maximum allowed by our Poisson forecast, and sometimes dropped below the minimum of our forecast, we could not construct a Poisson forecast that would be accepted by the N test for every target period (or even for $95\%$ of the target periods). Thus even for 5-year periods, a Poisson forecast is inappropriate when attempting to forecast all earthquakes (Supplement 3 demonstrates that the Poisson assumption seems justified for ill-defined mainshock-only forecasts). The next section presents a forecast with a non-Poissonian distribution of the number of events. 

\subsection{Negative Binomial Earthquake Number Distribution and Modified Likelihood Tests}
\label{sec:NBD}

We analyzed the distribution of the number of observed events in 5-year, non-overlapping intervals in the ANSS catalog of magnitudes $m \ge 4.95$ from 1932-2008 within the CSEP testing region. 
We compared the fit of the Poisson distribution to the empirical distribution with the fit of a negative binomial distribution (NBD). The discrete negative binomial probability mass function is defined by:
\be
p(k| \tau, \nu)= \frac{\Gamma(\tau+k)}{\Gamma(\tau) k!} \ \nu^{\tau} (1-\nu)^k
\ee
where $k=0,1,2,...,$ $\Gamma$ is the gamma function, $0 \leq \nu \leq 1$, and $\tau > 0$. There are many discrete distributions, but the NBD is simple, fits well and has been used before \citep{VereJones1970, Kagan1973, JacksonKagan1999}. The Akaike Information Criterion (AIC) favors the NBD ($AIC=130.4$) over the Poisson distribution ($AIC=246.1$). Using maximum likelihood, we obtained parameter estimates of the NBD ($[ \tau \approx 2.76, \nu \approx 0.08]$) and of the Poisson distribution ($\lambda \approx 29.9$). 


We created a NBD forecast, different from our Poisson forecast solely in the distribution of the number of events, summed over all space and magnitude bins (rates in individual bins remained the same). The two parameters of our NBD forecast can be calculated from the mean and variance of the distribution. For the mean, we preferred our earlier estimate (from section \ref{sec:ExpNum}) of the expected number of events, $N_{pred}=33.55$, based on the more recent, higher quality ANSS data from 1981 to 2009. To estimate the variance $Var(N_{obs}) \approx 368.1$, however, we used the longer data set from 1932. From these values, we obtained the NBD parameters ($[ \tau \approx 3.37, \nu \approx 0.09]$).

\begin{figure}
\centering
\includegraphics[draft=\IsDraft,width=\halfwidth,keepaspectratio=true,clip]{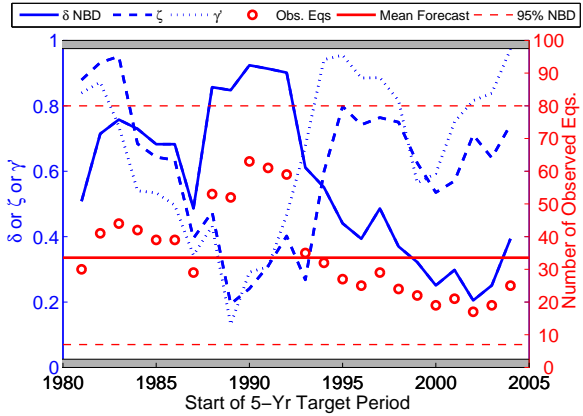}
\caption{\label{fig:relmRetroNBD}  Improved retrospective consistency tests of the 5-year negative-binomial forecast on past, overlapping 5-year periods with starting year as abscissa. Left (blue) ordinate: The negative binomial N-test statistic  $\delta$ (solid blue), measuring the agreement between the predicted and the observed number of total events; the spatial likelihood test statistic $\zeta$ (dashed blue), measuring the agreement between the expected and the observed spatial component of the likelihood, and the modified likelihood test statistic $\gamma'$ (dotted blue), measuring the agreement between the expected and the observed likelihood scores when normalized to the number of observed events. Right (red) ordinate: Number of observed events in each target period (red circles), along with the forecast's expected number of events (red solid line) and $95\%$ negative binomial confidence bounds (red dashed lines). 
}
\end{figure}

We repeated the retrospective N test on our new NBD forecast, the results of which are shown in Figure \ref{fig:relmRetroNBD}. Compared to Figure \ref{fig:relmRetro}, our new NBD forecast cannot be rejected for any target period, since the number of observed events is always well within the $95\%$ confidence bounds of our NBD distribution. 

We mentioned above that the $\gamma$ statistic conveyed little additional information over the $\delta$ statistic, because it depends too strongly on whether the observed number of events were within the $95\%$ confidence bounds of the number distribution. We therefore modified the L-test proposed by \citet{Schorlemmer-et-al2007} in two ways. For simulated likelihood scores, we used the same total number of events as were observed, conditioning the simulated likelihood scores on the observed number of events. In contrast, the L-test by \citet{Schorlemmer-et-al2007} simulated a random number of simulated events drawn from the Poisson distribution with mean equal to the total expected number of events. Thus each simulation in the original L-test has (potentially) a different number of events. This number is fixed in our modified L-test. We denote the probability that the observed likelihood value or less were simulated, conditional on the number of observed events, by $\gamma'$. The modified L-test tests the space and magnitude component of the forecast, but not the number of observed events. Results of the retrospective tests are shown in Figure \ref{fig:relmRetroNBD} (blue dotted curve). 

To solely test the spatial component of the forecast, we modified the L-test again by summing the forecast over all magnitude bins to eliminate the magnitude dimension. We then simulated likelihood values from this space-only forecast, conditioned on the number of observed events (see also \citep{Zechar-et-al2009b}). We denote the probability of simulating the observed spatial component of the likelihood score or a smaller value by $\zeta$, using \citet{Zechar-et-al2009b}'s notation. The results of this S-test are also shown in Figure \ref{fig:relmRetroNBD} (blue dashed curve). The model cannot be rejected by these new consistency tests, but we should expect this result since we used the same data set to build the model.  

In this section, we dropped the Poisson distribution in favor of the NBD to better forecast the number of earthquakes. 
This innocent, minor modification has major implications: (i) The sole time-independent stochastic point process is the Poisson process, and therefore any distribution but the Poisson distribution necessarily implies a time-dependent process. By acknowledging the NBD's better fit, we negate time-independent forecasting. (ii) We used a NBD for the total number of events, but in each individual bin remains specified as Poissonian. But summing the Poissonian rates over all bins cannot result in a NBD. 

How can we justify these inconsistencies? The short answer is: Because it is a quick and simple solution to a much deeper problem; An approximate solution that comes at the cost of a slight theoretical inconsistency. Moreover, the modified L-test we proposed no longer tests the overall number of observed events, and is thus to some extent decoupled from the NBD distribution. This is the approach we take in this article. However, the long answer is: We cannot justify it, and we must abandon time-independent earthquake forecasts. The Poisson process is an approximation that leads to great simplification, but the approximation is starting to catch up with us. We need time- and history-dependent process with memory, even for long-term forecasts (unless one forecasts so-called mainshocks (see Supplement 3), which can only be defined subjectively and retrospectively). Our ETAS branching model (section \ref{sec:ETAS}) is such a process, and in the future, we'd like to use it for long-term forecasts, too.


\section{Time-Independent Next-Day $m \geq 3.95$ Forecasts Based on Smoothed Seismicity}
\label{sec:fc2}

To compare our prospective next-day forecasts of the ETAS model, described next in section \ref{sec:ETAS}, to a simple yet informative and prospective reference model, we produced a time-independent next-day Poisson forecast based solely on smoothed seismicity. The forecast is identical to the five-year $m\geq 4.95$ forecast, except that we used $m_{min}=3.95$ as the minimum target magnitude and we calculated the expected number of events $N_{pred}$ per day for $m\geq 3.95$. There were $N_{pred}=0.177$ earthquakes $m\geq 3.95$ per day in the period from 1 January 1981 until 1 April 2009. 
We do not expect this forecast to perform well, since pulses of triggered earthquakes violate the time-independent nature of this forecast. 


\section{The Epidemic-Type Aftershock Sequences (ETAS) Model}
\label{sec:ETAS}
\subsection{Definition of the ETAS Model}

The Epidemic-Type Aftershock Sequence (ETAS) model \citep{Ogata1988, Ogata1998} is a stochastic spatio-temporal branching point-process model of seismicity. The model embodies the notion that every earthquake may trigger other earthquakes according to empirical probability distributions in time, space and magnitude. The triggering rate increases exponentially $~10^{\alpha m} $ with magnitude $m$ of the parent event and decays in time according to the Omori-Utsu law and in space according to a spatial decay function, e.g. a Gaussian or a power-law kernel. Each magnitude is identically and independently distributed according to the tapered Gutenberg-Richter distribution, independent of past seismicity. So-called aftershocks may thus be larger than the initiating shock. The ETAS model can model entire earthquake catalogs, rather than just aftershock sequences, because the total rate is determined by the sum of a time-independent background rate and the resulting cascades of triggered events. 

There are numerous flavors within the branching process family (see, e.g., \citet{KaganKnopoff1987, Kagan1991a, Felzer-et-al2002, HelmstetterSornette2002, Console-et-al2003b, Zhuang-et-al2004, HainzlOgata2005}). We used the particular formulation of \citet{Helmstetter-et-al2006}, albeit with some (minor) modifications. The total seismicity rate $\lambda(t,\vec{r},m)$ at time $t$, location $\vec{r}$ and for magnitude $m$ is the sum of a background rate $\mu_b(\vec{r})$, a spatially heterogeneous time-independent Poisson process, and the sum of individual contributions to the triggering potential from all prior earthquakes occurring at times $t_i<t$ with magnitudes $m_i$
\be
\lambda(t,\vec{r},m) = P_m(m) \left[ \mu_b(\vec{r}) + \sum_{t_i < t} \phi_{m_i} (\vec{r} - \vec{r_i}, t-t_i)  \right],
\label{eq:ETASrate}
\ee
where $P_m(m)$ is the space- and time-independent tapered Gutenberg-Richter distribution of magnitudes (\ref{tGR}). The triggering function $\phi_{m_i} (\vec{r}, t)$ describes the spatio-temporal triggering potential at a distance $\vec{r}$ and time $t$ after an earthquake of magnitude $m$
\be
\phi_m(\vec{r},t) = \rho(m) \psi(t) f(\vec{r}, m),
\label{eq:trigfun}
\ee
where $\rho(m)$ is the average number of earthquakes triggered by a quake of magnitude $m \ge m_{d}$
\be
\rho(m) = k 10^{\alpha (m-m_{d})}, 
\label{eq:rho}
\ee
the function $\psi(t)$ is the Omori-Utsu law normalized to 1
\be
\psi(t) = \frac{(p-1) c^{p-1}}{(t+c)^p}
\label{eq:Omori}
\ee
and $ f(\vec{r}, m)$ is a normalized spatial aftershock density at a distance $\vec{r}$ to the parent shock of magnitude $m$. The constant $m_d$ refers to a chosen threshold magnitude above which the model parameters are estimated, which may, in general, be different from the completeness magnitude $m_0$ and the target magnitude threshold $m_{min}$. 
We tested both a Gaussian and a power-law kernel, as described in section \ref{sec:SpaceAftershocks}. We fixed the c-value in Omori's law (\ref{eq:Omori}) to $0.035$ days (5 minutes). According to \citet{Helmstetter-et-al2006}, this parameter is not important as long as it is much smaller than the time window of the forecast (1 day). But sometimes, e.g. for the 1992 Landers earthquake, $c$ is greater than 1 day. Section \ref{sec:mc} discusses this problem in more detail, along with a solution by \citet{Helmstetter-et-al2006}.

We used the same tapered Gutenberg-Richter distribution (\ref{tGR}) as for the long-term forecast, with a b-value of $1.0$ and corner magnitude $m_{c}=8.0$ \citep{BirdKagan2004}. We used earthquakes $m \ge 2$ to forecast events $m_{min} \ge 3.95$ for the CSEP experiment (Supplement 5 presents parameter estimates for target threshold $m_{min}=2$). Since the earthquake catalog is not complete down to $m_{d}=2$ after large earthquakes, we corrected for this effect (see section \ref{sec:mc}), as well as for any time-independent spatial incompleteness. 

The background rate $\mu_b$ is given by 
\be
\mu_b(\vec{r}) = \mu_s \mu_0(\vec{r})
\label{eq:bg}
\ee
where $\mu_0(\vec{r})$ is equal to our favorite spatial, time-independent model normalized to one (model 21 in Table \ref{tab:SmoOpt}), so that the parameter $\mu_s$ represents the expected number of $m \ge m_{min}$ background events per day. Since we used the same spatial model for the time-independent $m \ge 3.95$ forecasts (section \ref{sec:fc2}), we can evaluate directly the improvement of the ETAS forecasts over a static model.

\subsection{ETAS Forecasts for 1-Day Bins}

The ETAS model is defined via its conditional intensity or hazard rate (\ref{eq:ETASrate}), which is the instantaneous probability of an event \citep{DaleyVereJones2004}. To forecast the number of earthquakes over a finite period, one cannot directly use the conditional intensity, since intervening quakes may change the rate. One solution to forecast over a 1-day period is to simulate earthquakes according to the conditional intensity at the beginning of the 1-day bin, and then to obtain a mean forecast by averaging over all simulations. Such simulations are computationally intensive, and it is not clear whether the $10,000$ simulations commonly used to produce spatial-temporal forecasts are adequately sampling the possibilities. 

A simpler solution is the following. \citet{HelmstetterSornette2003d} showed that, for synthetic ETAS catalogs, the use of $N_p=\int_{t_p}^{t_p+T} \lambda(t) dt$ to predict the number of quakes between $t_p$ and $t_p+T$ underestimates the number of actually occurred earthquakes by an approximate constant factor, independent of the number of future events. 
We can use the ETAS model (\ref{eq:ETASrate}) to forecast the number of events of the next day, but with effective parameters $k, \alpha$ and $\mu_s$, which are different from the original parameters of the ETAS model \citep{Helmstetter-et-al2006}. Instead of using the likelihood of the ETAS model, we estimated parameters by maximizing the likelihood of the next-day forecasts using a Poisson likelihood defined in section \ref{sec:LL}. These effective parameters depend on the forecast horizon and are difficult to compare to the original ETAS parameters. 

\subsection{Correcting for Spatial and Temporal Magnitude Incompleteness}
\label{sec:mc}

The ANSS catalog is not complete down to magnitude $m=2$ everywhere in California. To correct for the spatial incompleteness, we used the same completeness magnitude estimate as for our time-independent forecast, as described in section \ref{sec:mcSpace} and shown in Figure \ref{fig:mc}. However, for our time-dependent forecasts, the temporary incompleteness of catalogs after large earthquakes also becomes important. \citep{Kagan2004, Helmstetter-et-al2006, Peng-et-al2007, Lennartz-et-al2008} showed that the completeness magnitude after large earthquakes can temporarily increase by several magnitudes. One effect is that the forecast overestimates the observed rate because of missing events. Another effect is that secondary triggering is underestimated because early undetected aftershocks contribute to the rate. We followed the solutions proposed by \citet{Helmstetter-et-al2006} to correct both effects. 

We estimated the completeness magnitude $m_0(t,m)$ as a function of time $t$ (in days) after an event of magnitude $m$ using:
\be
m_0(t,m) = m - 4.5 - 0.75 \log_{10} (t)
\label{eq:mct}
\ee
and 
\be
m_0(t,m) \ge 2
\ee
Figure 6 of \citet{Helmstetter-et-al2006} 
illustrates this relation for sequences triggered by the 1992 $M_W7.3$ Landers, the 1994 $M_W6.7$ Northridge and the 1999 $M_W7.1$ Hector Mine earthquakes. 


We used expression (\ref{eq:mct}) to estimate the detection threshold at the time of each earthquake. This time-dependent threshold is usually larger than the regular spatially-varying threshold for earthquakes that occur shortly after large $m>5$ earthquakes. We selected only earthquakes with $m>m_0$ to estimate the seismicity rate (\ref{eq:ETASrate}) and to calculate the likelihood of the forecasts. 

We also corrected the forecasts for the second effect, the missing contribution from undetected aftershocks $m_d < m < m_0(t)$ to the observed seismicity rate: We added a contribution to the rate $\rho(m)$ due to detected earthquakes $m>m_0(t)$ (\ref{eq:rho})
\be
\rho^*(m) = \rho(m) + \frac{K b }{b-\alpha} 10^{b(m_0(t)-m_d)} \left[ 1-10^{-(b-\alpha)(m_0(t)-m_d)}\right] .
\label{eq:rhostar}
\ee
where $m_0(t)$ is the detection threshold at the time $t$ of the earthquake, estimated by (\ref{eq:mct}), due to the effect of all prior $m>5$ earthquakes. The second contribution corresponds to the effect of all earthquakes with $m_d < m < m_0(t)$ that occur on average for each detected earthquake. This contribution is of the same order as the contribution from the observed events for reasonable parameter values, because a large fraction of aftershocks are secondary aftershocks and because small earthquakes collectively trigger an equal amount of aftershocks as larger ones if $\alpha=b$.

\subsection{Modeling the Spatial Distribution of Aftershocks}
\label{sec:SpaceAftershocks}

We tested different choices for the spatial kernel $K_{d(m)}(\vec{r}, m)$, which models the aftershock density at a distance $r$ from a parent shock of magnitude $m$. As in section \ref{sec:AK}, we compared a power-law function (\ref{eq:Kpl}) with a Gaussian kernel (\ref{eq:Kgauss}). 
The spatial regularization distance $d(m)$, which replaces the adaptive bandwidth $d_i$ in the kernels (\ref{eq:Kpl}) and (\ref{eq:Kgauss}), accounts for the finite rupture size and for location errors. We assumed that $d(m)$ is given by
\be
d(m) = 0.5 + f_d \cdot 0.01 \cdot 10^{0.5m} \rm{km},
\label{eq:dm}
\ee
where the first term on the right-hand-side accounts for location accuracy and the second term represents the aftershock zone length of an earthquake of magnitude $m$. The parameter $f_d$ is estimated by optimizing the forecasts and, in the case of the Gaussian kernel (\ref{eq:Kgauss}), should be close to one if the aftershock zone size is isotropic and $67\%$ (one standard deviation) of triggered events fall within one rupture length. The rupture length estimate (\ref{eq:dm}) is similar to the expression by \citet{WellsCoppersmith1994}.

The choice of the exponent $1.5$ in the power-law kernel (\ref{eq:Kpl}) is motivated by recent studies \citep{Ogata2004, Console-et-al2003b, Zhuang-et-al2004}, who inverted this parameter in earthquake catalogs by maximizing the likelihood of the ETAS model, and who all found an exponent close to $1.5$. This value also conveniently translates into an analytically integrable form. It predicts that the aftershock density decays with distance $r$ from the parent shock as $1/r^3$ in the far field, proportional to the static stress change.

Large $m>5.5$ earthquakes with a rupture length larger than the grid cell size of $0.1^\circ$ are rarely followed by isotropic aftershock distributions. We therefore used a more complex, anisotropic kernel for these events, as done previously by \citet{WiemerKatsumata1999}, \citet{Wiemer2000} and \citet{Helmstetter-et-al2006}. We smoothed the locations of early, nearby aftershocks to estimate the main shock fault plane and other active faults in the immediate vicinity. We computed the distribution of later aftershocks of large $m \ge 5.5$ quakes by smoothing the locations of early aftershocks using
\be
K_{d}(\vec{r},m) = \sum_{i=1}^N K_d(|\vec{r} - \vec{r_i}|,m_i),
\label{eq:fea}
\ee
where the sum is over the main shock and all earthquakes that occur within a distance $D_{aft}(m)$ before the issue time $t_p$ of the forecast and not after some time $T_{aft}$ after the main shock. We usually took $D_{aft}=0.02 \times 10^{0.5m}$ km (approximately two rupture lengths) and $T_{aft}=2$ days but tested other values in section \ref{sec:PV}. 
The kernel $K_d(\vec{r}, m)$ used to smooth the locations of early aftershocks is either a power-law (\ref{eq:Kpl}) or a Gaussian distribution (\ref{eq:Kgauss}), with an aftershock zone length given by (\ref{eq:dm}) for the main shocks, but by $d=2$ km for the aftershocks. Figure 7 of \citet{Helmstetter-et-al2006} compares the Gaussian and the power-law kernel on the 1992 $M_W7.3$ Landers earthquake sequence. 

Smoothing the locations of early aftershocks to forecast the spatial distribution of later aftershocks is a fast and completely automatic method to estimate the main shock rupture plane along which aftershocks tend to cluster. In section \ref{sec:PV} we tested several values of the method's parameters. 

\subsection{Definition of the Likelihood and Estimation of the ETAS Model Parameters}
\label{sec:LL}

We estimated the parameters of the ETAS model by maximizing the likelihood of the data. We inverted for five parameters: $p$ (the exponent in Omori's law (\ref{eq:Omori})), $k$ and $\alpha$ (characterizing the productivity (\ref{eq:rho})), $\mu_s$ (the number of background events per day (\ref{eq:bg})), and $f_d$ (the size of the aftershock zone (\ref{eq:dm})). 

As already stated, we did not maximize the likelihood function of the ETAS model. Rather, we estimated effective parameters by maximizing the cumulative likelihood of the next-day forecasts using a Poisson distribution (\ref{eq:Poi}). 
The log-likelihood of the forecasts is the sum of log-likelihoods in each space-time-magnitude bin indexed by $(i_t, i_x,i_y,i_m)$:
\be
LL = \sum_{i_t=1}^{N_t}  \sum_{i_x=1}^{N_x}  \sum_{i_y=1}^{N_y}  \sum_{i_m=1}^{N_m} \log p\left( N_p(i_t, i_x,i_y,i_m), n \right)
\label{eq:LL}
\ee
where $n$ is the number of observed events in the bin $(i_t, i_x,i_y,i_m)$ and the probability $p(\cdot, n)$ is a Poisson distribution with a rate given by the expected number of events $N_p(i_t, i_x,i_y,i_m)$ (the forecast), which in turn is the integral over each space-time-magnitude bin of the predicted seismicity rate $\lambda(\vec{r}, t, m)$
\be
N_p(i_t, i_x,i_y,i_m) = \int_{i_t} \int_{i_x} \int_{i_y} \int_{i_m} \lambda(\vec{r}, t, m) \ dm \ dy \ dx \ dt .
\label{eq:Np}
\ee
The rules of the CSEP one-day forecast group determine the bin size: 1 day in time, $0.1$ degrees longitude and latitude in space, and 0.1 units of magnitude \citep{SchorlemmerGerstenberger2007}. 

The log-likelihood (\ref{eq:LL}) can be simplified by substituting the Poisson distribution (\ref{eq:Poi}) and noting that the sum over the space-magnitude bins need only be carried out over those bins in which earthquakes occurred
\be
LL = \sum_{i_t} \left[ -N_p(i_t) + \sum_{i_x=1}  \sum_{i_y=1}  \sum_{i_m=1} n \ \log \left[ N_p (i_t,i_x,i_y,i_m) \right] - \log(n!) \right]
\label{eq:LLs}
\ee
where $N_p(i_t)$ is the total expected number of events in the time bin $i_t$ between $t(i_t)$ and $t(i_t+T)$ 
\be
N_p(i_t) = \mu_s + \int_{i_t} \sum_{t_i<t(i_t)} f_i \rho(m_i) \psi (t_p - t_i) \ dt .
\label{eq:Npti}
\ee
The factor $f_i$ in (\ref{eq:Npti}) is the integral of the spatial kernel $f_i(\vec{r} - \vec{r_i})$ over the grid, which is smaller than 1 due to the finite grid size. 

We maximized the log-likelihood (\ref{eq:LL}) using a simplex algorithm (\citet{Press-et-al1992}, p. 402) and using earthquakes above various $m_{min}$ from 1 January 1986 until 1 April 2009 in the CSEP testing region to test the forecasts. To calculate the seismicity rate (\ref{eq:ETASrate}), however, we took into account earthquakes $m \ge 2$ since 1 January 1981 that occurred within CSEP collection region. We tested the effect of variations of the spatial aftershock kernel, and different target magnitude thresholds. 

To quantify the performance of the short-term forecasts with respect to the time-independent forecast, we used, as before, the probability gain per earthquake (\ref{eq:G}) of the ETAS model likelihood with respect to the time-independent likelihood. 
The time-independent model is our favorite background density $\mu (\vec{r})$ (model 21 in Table \ref{tab:SmoOpt}), which also serves as the background model of the ETAS model, but we normalized $\mu (\vec{r})$ so that the total number of expected target events equalled the observed number. 

Since the background model was built from the same data that the time-dependent model was tested on, the probability gain solely measures the relative increase of the spatio-temporal aspect of the ETAS forecast over the time-independent model. In a truly prospective mode, the actual likelihood values may be smaller, although the improvement over a time-independent forecast may remain high. 

By maximizing the likelihood of the Poisson forecasts, we assumed that the Poisson distribution in each space-time-magnitude bin is a first-order approximation to the actual, model-dependent distribution. The actual likelihood function of the ETAS model is known but the predictive next-day likelihood function must be simulated and can deviate substantially from the Poisson distribution. However, rather than performing computationally intensive simulations, we used the current model to estimate the mean rate in each bin. Extensions beyond the Poisson distribution will be left for the future.

\subsection{Estimated Parameter Values}
\label{sec:PV}

The parameter estimation program was modified from the one written by \citet{Helmstetter-et-al2006}: We extended the region to the CSEP collection region of California, and we fixed some issues in the code. We tested different versions of the ETAS model (various spatial kernels, different learning and target magnitude thresholds, different parameter values of the early-aftershock-kernel smoothing procedure, etc.). Table \ref{tab:optETASm4}  presents the results. For brevity, we discuss the performance of the optimization algorithm in Supplement 4. 

The exponent $\alpha$  in the productivity law (\ref{eq:rho}) measures the relative importance of small versus large earthquakes for the triggering budget \citep{Helmstetter2003, Felzer-et-al2004, Helmstetter-et-al2005, ChristophersenSmith2008, Hainzl-et-al2008}. \citet{Felzer-et-al2004, Helmstetter-et-al2005} found that $\alpha$ is close to or equal to one by fitting the number of aftershocks as a function of main shock magnitude. Since the number of small earthquakes increases exponentially with decreasing magnitude, their collective ability to trigger earthquakes equals that of the large earthquakes \citep{Helmstetter2003}. As a consequence, small, undetected earthquakes have a significant, time-dependent impact on the observed seismicity budget and the failure to model their effect causes parameter bias \citep{SornetteWerner2005a, SornetteWerner2005b, SaichevSornette2005b, SaichevSornette2006, Werner2007, Zhuang-et-al2008}. We found $\alpha = 0.8 \pm 0.1$ for all models for targets $m \ge 3.95$. Because of the strong negative correlation between $\alpha$ and $K$, it is possible that $\alpha=1$ is within the uncertainties of the estimates. The estimate of $\alpha$ decreases when the aftershock kernel is purely isotropic and does not include the smoothing of early aftershocks, consistent with the simulations by \citet{Hainzl-et-al2008}.  \citet{Helmstetter-et-al2006} estimated $\alpha \approx 0.43$ using essentially the same optimization procedure that we used, but for the target magnitude range $m \ge 2$. Supplement 5 lists the estimated parameters for the $m \ge 2$ range: Here, our estimates are almost identical to those of \citet{Helmstetter-et-al2006}. 

The exponent $p$ in Omori's law was relatively high $p \approx 1.27$, although typical estimates use smaller magnitudes and maximize the likelihood of Omori's law, not of the Poisson forecasts. 

The background rate was also fairly stable across the different models, at $\mu_s \approx 0.08$, which was about $44\%$ of the average daily number of earthquakes $m \ge 3.95$ over the entire period. The estimated proportion of triggered events is thus $56\%$. However, this is a lower bound on the actual proportion because events below $m_0(t)$ were not counted, and any estimate of the fraction of triggered events depends on the magnitude threshold \citep{SornetteWerner2005b}. The background rate of models with the Gaussian kernel was slightly larger than of models with the power-law kernel, because the latter has much farther reach than the Gaussian kernel. 

The parameter $f_d$ is a measure of the size of the aftershock zone. 
As we explained above, for the Gaussian spatial kernel (\ref{eq:Kgauss}), the length $d(m)$ (\ref{eq:dm}) equals the standard deviation of the Gaussian, so that $f_d$ 
should be about $1$. This parameter can be used as a sanity check of the optimization. We found reasonable values for $f_d$ between $0.5$ and $0.9$ in the case of the Gaussian kernel. 

Using the power-law kernel, $f_d$ was very small: between $0.08$ and $0.25$. When we did not smooth the locations of early aftershocks to forecast later ones, the value increased to a more reasonable $0.25$. 
Nevertheless, these values are difficult to interpret. It suggests using the Gaussian kernel since the fit is better understood, despite the marginally better performance of the power-law kernel. 


A lower probability gain per earthquake was obtained, for both spatial kernels, when we did not smooth the locations of early aftershocks to forecast the locations of later aftershocks: For example, model 1, which used the anisotropic smoothing method for events $m>5.5$, scored $G=6.19$, while model 7, which used the isotropic kernel, obtained $G=6.04$. The gains of the smoothing method are particularly strong during the early days in an aftershock sequence, as expected. We calculated the average daily log-likelihood ratio for the first 10 days after all 18 $m>6$ events in the target period and found that model 1 outperformed model 7 by an average daily log-likelihood ratio of 0.28. During the aftershock sequences, the smoothing method thus works better than the isotropic kernel. However, the gains are somewhat diluted because the parameters that use the isotropic kernel tend to give better forecasts on the days of the actual $m>6$ events. There seems to be a subtle trade-off effect here. The benefits of the smoothing technique did not change much when we varied the temporal window for smoothing from 1 to 3 days nor when we lowered the magnitude threshold for selecting events from $m=6$ to $m=5$ (see models 5 through 9). 

We tested whether adding the contribution $\rho*$ (\ref{eq:rhostar}) of undetected events $m_d<m<m_0(t)$ increases the likelihood values: This was not the case over the entire period. However, certain individual sequences were much better fit using this correction, e.g. the 1992 $M_W7.3$ Landers earthquake sequence (not shown). There seems to be a subtle trade-off effect between modeling sequences and performing well on the days of the actual mainshocks.

\subsection{Observed and Predicted Number of Events}

\begin{figure}
\centering
\includegraphics[draft=\IsDraft,width=\threequarterwidth,keepaspectratio=true,clip]{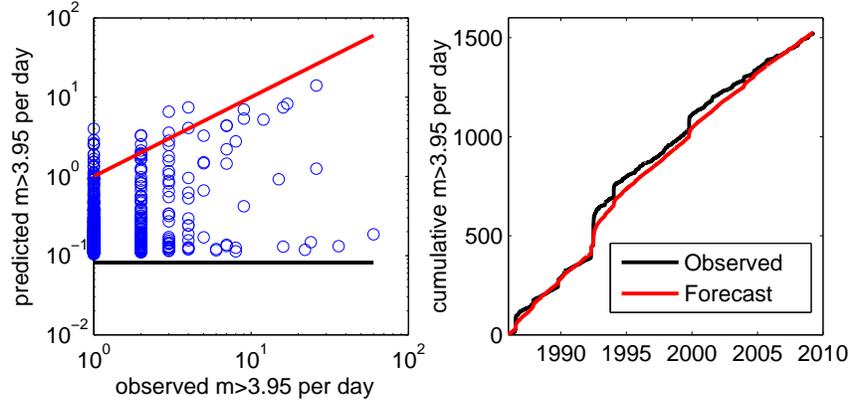}
\caption{\label{fig:obspredcum} Left: Comparison of the daily expected number of earthquakes with observed numbers (blue circles). The solid red line corresponds to a perfect match. The solid black line represents the background rate of the ETAS model. Right: Comparison of cumulative observed number of $m \ge 3.95$ events and the cumulative daily ETAS forecast. 
}
\end{figure}

Figure \ref{fig:obspredcum} compares the observed (target) earthquakes and the predicted number of events. The left panel compares the daily forecasts against the daily observations. Three groups of days can be identified. Firstly, there are days on which the forecasts are low (close to the background rate) and few events occur that match the forecast. The second group corresponds to low forecasts near the ETAS background rate, but to many observed events that are inconsistent with the model. On those days, large earthquakes and triggered sequences occurred without foreshocks on previous days that could have helped the forecast. But the model's performance was worse than it could be: The CSEP rules for the one-day forecast group allow an update of the forecast only once every 24 hours. The full potential of the ETAS model can only be realized if updates are allowed after every earthquake. The third group close to the red line (marking the perfect match) shows the potential: These are days of active sequences that started the day before, the information of which allowed the model to reasonably forecast seismicity. 

The cumulative forecasts (right panel in Figure \ref{fig:obspredcum}) followed the observed number of events relatively well, including the aftershock sequences. But the forecast systematically underestimated individual sequences, e.g. the period after the 1992 $M_W7.3$ Landers earthquake. We expected this underprediction as the model cannot update after each event. At the same time, the total cumulative number of events was matched well at the end of the entire period. This is because the likelihood penalizes for mismatches in the total number of events. But since the days during which large quakes and their aftershocks occur cannot be accurately forecast because of the updating rules, the parameters are biased: The background rate is overestimated to match the total number of events. These complications are a result of the one-day forecasts that are unnecessary for many model families. At the same time, the ETAS model will remain a poor predictor of strong events if no foreshocks raise the forecast.

\begin{figure}
\centering
\includegraphics[draft=\IsDraft,width=\mediumwidth,keepaspectratio=true,clip]{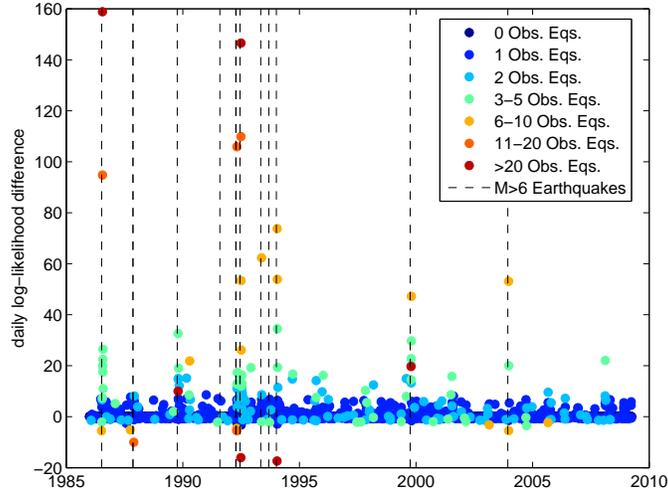}
\caption{\label{fig:DailyLL} Daily log-likelihood ratios between the ETAS model and the time-independent forecast (filled circles). The circles are color-coded by the number of events that occurred on each day. Also shown by vertical dashed lines are all $m>6$ events during this period.  
}
\end{figure}

How did the ETAS model perform against the time-independent forecast over the course of the entire period? Figure \ref{fig:DailyLL} shows the daily log-likelihood ratios between the ETAS model and the time-independent forecast. The largest ratios occurred on or just after large $m>6$ earthquakes, which we marked by vertical dashed lines. The daily log-likelihood ratios are also color-coded by the number of observed events on each day. The performance of the two models was very similar for days on which no earthquakes occur (dark blue filled circles are obscured by overlying markers). The differences increased with the number of observed events. For instance, on the day of the 1999 Hector Mine earthquake, many earthquakes occurred, and the ETAS model strongly outperformed the Poisson model forecast. Several small earthquakes occurred before this large event, thereby locally increasing the ETAS forecast with respect to the Poisson model (see also Figure \ref{fig:M6} discussed below). 

On days when the ETAS rate decayed to its background level, the time-independent forecast had a higher rate, so that if an earthquake occurred, the Poisson model beat the ETAS forecast. But the likelihood ratio on those days was never very large, since the background rate of the ETAS model is not much smaller than the Poisson rate. Therefore, the ETAS model is to some extent guarded against surprise events by its background rate, while making vastly better forecasts during active sequences. 

Figure \ref{fig:DailyLL} shows that the ETAS model performed significantly better than the Poisson model after all large $m>6$ earthquakes. Additionally, it performed better on the days of large events that were preceded by foreshocks. We discuss a few individual earthquake sequences below.

\subsection{Observed and Modeled Temporal Distribution of Aftershocks}

\begin{figure}
\centering
\includegraphics[draft=\IsDraft,width=\mediumwidth,keepaspectratio=true,clip]{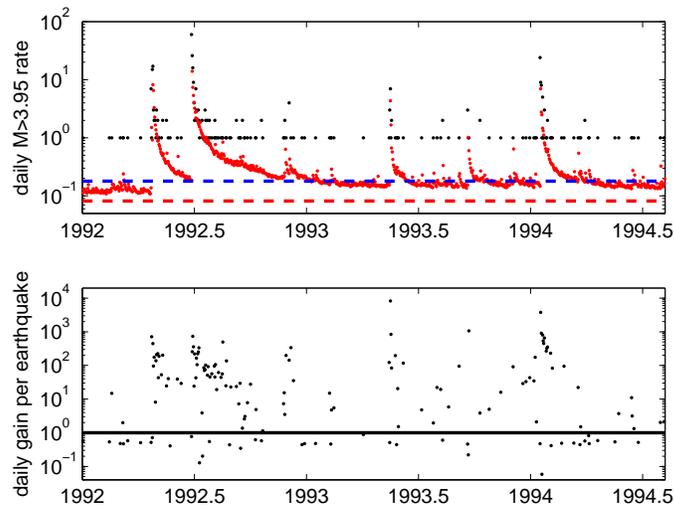}
\caption{\label{fig:RatesG} Top panel: Comparison of observed events per day (black dots) with the ETAS forecast (red dots). The blue dashed horizontal line represents the time-independent forecast calculated from 1986 until 2009, while the red dashed line shows the ETAS background rate. Bottom panel: Daily probability gains per earthquake (black dots) and reference level of no gain, $G=1$, (solid black line).
}
\end{figure}

Figure \ref{fig:RatesG} compares the observed number of events with ETAS forecasts from 1992 until August 1994, along with the daily probability gain per quake (\ref{eq:G}), which normalizes the exponent of the likelihood ratio by the number of observed events per day. Also shown are the background rate of the ETAS model (dashed red line) and the rate of the time-independent forecast (dashed blue line). The figure shows how the ETAS forecast tracked the observed seismicity during the aftershock sequences of this particularly active period. The largest gains correspond to the most active days. Individual gains were as large as $G \sim 10^4$ per day per earthquake. 

\begin{figure}
\centering
\includegraphics[draft=\IsDraft,width=\threequarterwidth,keepaspectratio=true,clip]{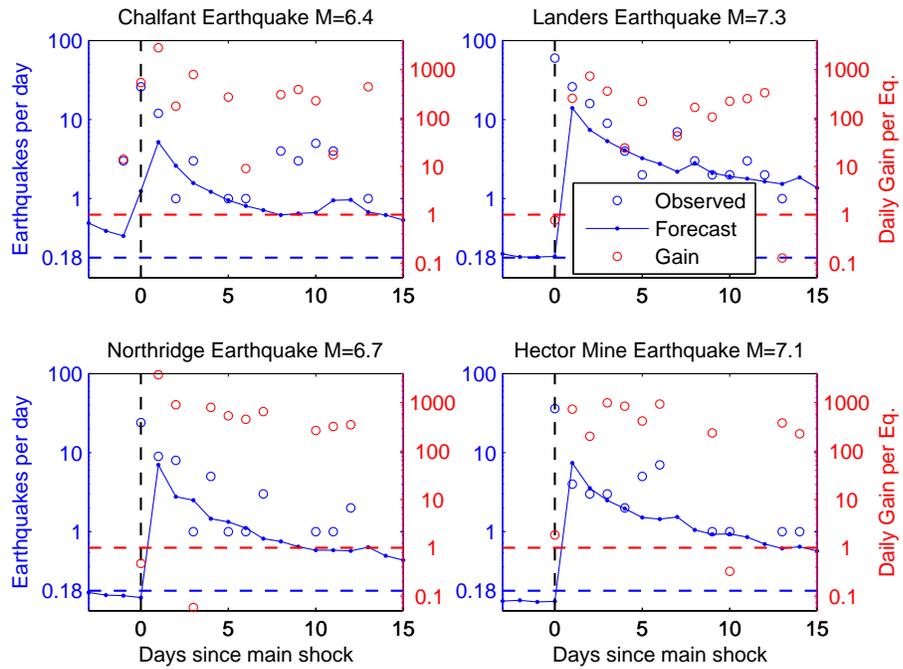}
\caption{\label{fig:M6} Retrospective daily forecasts of the ETAS model (blue curve) during four triggered earthquake sequences compared with the observed number of events (blue circles) and the time-independent forecast (horizontal dashed blue line). On the right (red) ordinate axis, the panels show the daily probability gain per earthquake of the ETAS forecast over the time-independent forecast (red circles). The red dashed line marks the no-gain line. Vertical black dashed lines mark the day of the four large $m>6$ earthquakes. 
}
\end{figure}

To get an even more detailed picture of the ETAS model's performance in individual aftershock sequences, we compared the forecast and the observations just before and after 4 major earthquakes in Figure \ref{fig:M6}. The figure shows the ETAS model forecast, integrated over space and magnitude bins, along with the time-independent forecast (dashed blue line). Also shown are the actually observed number of events on each day, along with the daily probability gains per earthquake on the red axis scale on the right. The four earthquakes illustrate two successes and two failures of the ETAS model compared to the time-independent forecast. The 1986 $M_W6.4$ Chalfant earthquake occurred when the ETAS forecast was higher than the Poisson forecast, as even on the day before the event the ETAS model outperformed the time-independent forecast. On the day of the event, the model underestimated the number of events because when the forecast was made, the large main shock had not yet occurred. In the following days, the ETAS model forecast the number of events better than the Poisson forecast, as measured by the large gains (red circles). The 1992 $M_W7.3$ Landers earthquake sequence shows similar properties, except that the ETAS model rate was slightly below the Poisson forecast on the day of the shock so that the gain was slightly below 1. The 1994 $M_W6.7$ Northridge earthquake was also not well forecast by the ETAS model, its rate having sunk below the Poisson forecast. Nevertheless, the gains during the aftershock sequence are significant. The 1999 $M_W7.1$ Hector Mine earthquake, on the other hand, is a success story for the ETAS model. Although its total rate, summed over the region, is smaller than the Poisson forecast (as can be seen in Figure  \ref{fig:M6}), a few small quakes locally increased the ETAS rate above the Poisson forecast, so that the ETAS model outperformed the Poisson model on the day of the Hector Mine earthquake. 

Figure \ref{fig:M6} can also be used to judge the fit of the ETAS model aftershock forecasts with the actual observed events. The fluctuations in the number of observed aftershocks are clearly larger than the (mean) ETAS model forecast. Even $95\%$ confidence bounds of the Poisson distribution around the mean forecast cannot enclose the observations. So while the likelihood ratio and the probability gain registered an immense improvement of the ETAS model over the time-independent model, we did not test whether the observations are consistent with the ETAS model, as we did for our long-term forecast in section \ref{sec:retro}. If we were to apply daily the current CSEP number test \citep{Schorlemmer-et-al2007}, which assumes a Poisson uncertainty in the number of events, the model would be rejected on many days. However, at this point it is clear that the original RELM tests are inappropriate for one-day forecasts. Rather, a first step in the right direction would be a model-dependent, simulated likelihood distribution against which the observations are counted \citep{WernerSornette2008a}. Because of the computational complexity, however, we need to leave this extension to future work. 

\subsection{Observed and Modeled Spatio-Temporal Distribution of Aftershocks}

\begin{figure}
\centering
\includegraphics[draft=\IsDraft,width=\threequarterwidth,keepaspectratio=true,clip]{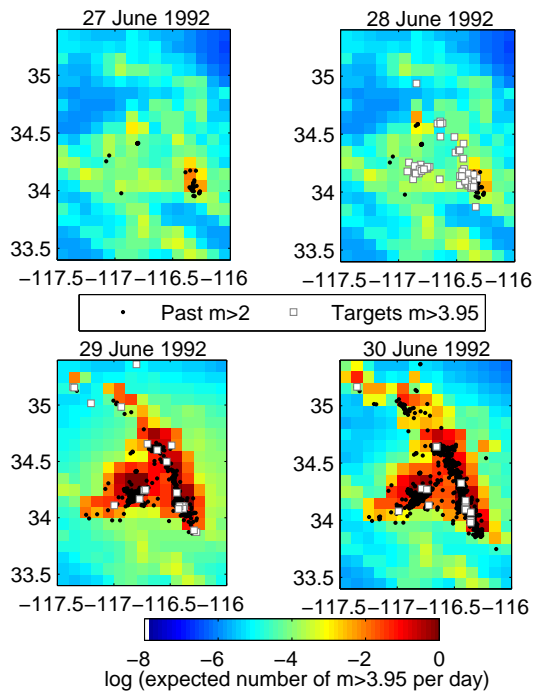}
\caption{\label{fig:L1} ETAS model forecast using a Gaussian spatial kernel and early-aftershock smoothing for the region surrounding the 1992 $M_W7.3$ Landers earthquake epicenter for the day before the main shock (27 June), of the main shock (28 June), and the two subsequent days (29-30 June). 
Also shown are $m \ge 2$ that occurred in the week prior to the day (small black dots) and the target $m \ge 3.95$ events that occurred on the forecast day (white squares with grey lining). 
}
\end{figure}

Figure \ref{fig:L1} compares the observed events with the spatio-temporal forecast around the 1992 $M_W7.3$ Landers earthquake. The panels show the forecast on the day before the main shock (27 June 1992), on the day of the main shock (28 June 1992) and on two subsequent days (29-30 June 1992). Also shown are all small events $m \ge 2$ that occurred in the previous week and contributed significantly to the ETAS model forecast, along with the target $m \ge 3.95$ earthquakes of each day. We observed large differences between the Gaussian and power-law kernels on the days after the main shock: The power-law kernel has a larger and smoother spatial extent than the Gaussian kernel (see Figure 7 of \citet{Helmstetter-et-al2007}). The power-law kernel forecasts a high rate in many bins in which no events occur. But on the other hand, it better predicted remote targets in bins unaffected by a Gaussian forecast. We preferred the Gaussian kernel because the estimates for the parameter $f_d$ seemed more reasonable (see section \ref{sec:PV}).

\section{ Time-Dependent Next-Day $m \geq 3.95$ Forecasts Based on the ETAS Model}
\label{sec:TD}

\begin{figure}
\centering
\includegraphics[draft=\IsDraft,width=\fullwidth,keepaspectratio=true,clip]{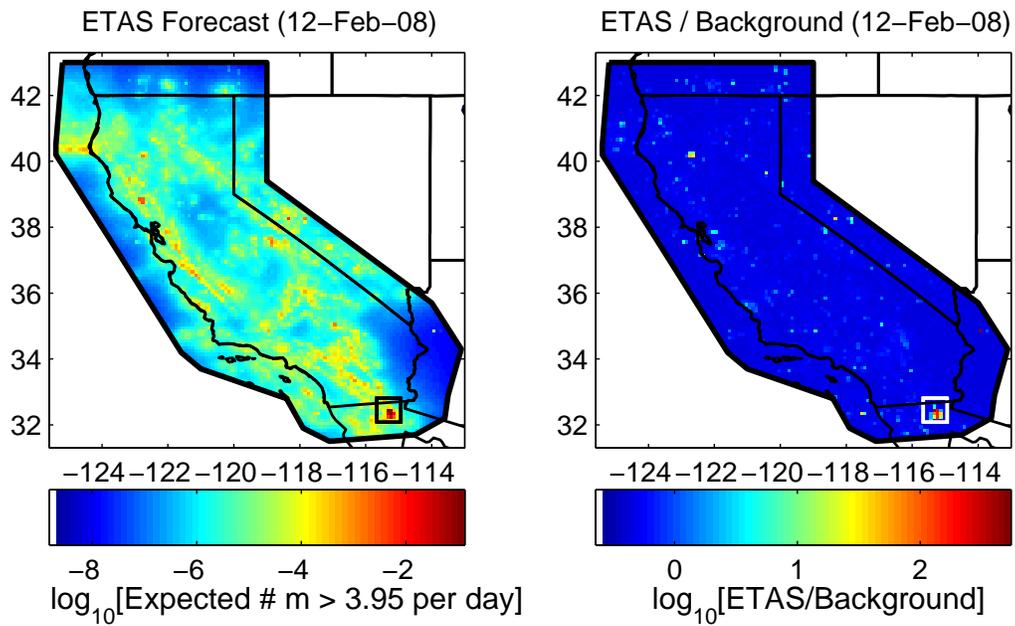}
\caption{\label{fig:cf}  Left: ETAS model forecast for 12 February 2008 along with the four $m \ge 3.95$ observed events on that day (black squares, overlapping). Right: Comparison of the ETAS model forecast and the time-independent smoothed seismicity forecast, along with observed events that day (white squares, overlapping). Because of several earlier quakes, the ETAS model predicts the occurrence of the four observed events on that day much better than the Poisson model. 
}
\end{figure}

An example of an ETAS model forecast (using model 1 in Table \ref{tab:optETASm4}) is shown in Figure \ref{fig:cf} along with a comparison to the time-independent $m \ge 3.95$ forecast described in section \ref{sec:fc2}. The plot on the left shows the expected number of earthquakes of magnitudes $m \ge 3.95$ in each cell on February 12, 2008. On that day, four earthquakes $m \ge 3.95$ occurred in roughly the same location in Baja California, Mexico, as part of a series of about ten swarm-like earthquakes during a two-week period. Because on previous days several earthquakes occurred, the ETAS model rate is locally higher than the time-independent forecast by a factor of almost $1,000$. Other differences between the ETAS model and the background model are less pronounced and mainly due to small, recent events that increased the ETAS model rate. 

For our time-dependent forecasts, we used the same magnitude distribution as for the time-independent forecast described in section \ref{sec:MD}, including the same corner magnitude $m_c=8.0$. 


\section{Discussion and Conclusions}

We presented two models, modified from  \citet{Helmstetter-et-al2006,Helmstetter-et-al2007}, for estimating the probabilities of future earthquakes in California. 
The time-independent model uses an optimized adaptive kernel to smooth the locations of small $m \ge 2$ declustered earthquakes. The model corrects for spatial magnitude incompleteness and assumes a tapered Gutenberg-Richter distribution with corner magnitude equal to $8.0$. While we corrected and improved the procedure for estimating the completeness magnitude, we are still unsatisfied with its performance. In the future, we'd like to use a more robust method, such as the one by \citet{SchorlemmerWoessner2008}. Another area for improvement involves the use of anisotropic kernels to smooth seismicity, although our method to smooth the locations of early aftershocks to forecast those of later ones seems to perform quite well. Finally, the earthquake catalog we used is relatively short (28 years) compared to average recurrence times of very large earthquakes and the crust's memory. In the future, we'd like to find a trade-off between high-quality recent data and lower quality older data. 

Our original long-term forecast was cast as a time-independent Poisson process. However, the distribution of the number of earthquakes in any finite time period is not a Poisson distribution and better described by a negative binomial distribution \citep{Kagan1973, JacksonKagan1999, Schorlemmer-et-al2009} or perhaps a heavy-tailed distribution \citep{SaichevSornette2006c}. Retrospective tests showed that the Poisson assumption is violated more often than expected even for 5-year target periods. We therefore modified our time-independent forecast such that the distribution of the number of forecast events was a negative binomial distribution, with mean equal to our Poisson forecast but a variance estimated from past data. The modified long-term forecast passed retrospective number consistency tests. We also modified the likelihood consistency test by conditioning the simulated likelihood values on the number of observed events. We thereby increased the test's spatial resolution and made it less sensitive to the number of observed events. 

While we modified the number distribution of our long-term model, we did not specify the variances in each individual space-magnitude bin. This inconsistency can be solved by abandoning time-independent forecasts and replacing the Poisson process by a time-dependent process of earthquake clustering. Therefore, calculating more realistic 5-year forecasts will involve simulations of (or approximations to the distributions of) a branching process. 

Our time-dependent model is such a branching model, but here we solely used it for average next-day forecasts. To compete in CSEP's one-day forecast group, we specified next-day forecasts as Poisson forecasts that change daily. This is a rough approximation \citep{WernerSornette2008a}, since the rates vary significantly within one day and bins are not spatially independent. But the extension to a simulated model-dependent number and likelihood distribution for each bin will be left for the future. 

The time-dependent next-day forecasts are based on the ETAS model, the parameters of which we optimized on retrospective forecasts using the assumed Poisson distribution. Other features that distinguish our implementation from other ETAS flavors include (i) a method to correct for spatial and temporal magnitude incompleteness, (ii) smoothing the locations of early aftershocks to forecast later ones, and (iii) using small $m \ge 2$ earthquakes to forecast larger $m \ge 3.95$ events. The ETAS model forecasts outperformed the time-independent forecast with a probability gain per earthquake of about $6$. 

We expect our models to perform well, based on \citep{Schorlemmer-et-al2009}'s report on the success of the time-independent model by \citep{Helmstetter-et-al2007} in the RELM experiment. However, with increasing time span and the occurrence of a very large event, smoother models based on tectonic, geological and geodetic data might work better. Determining the reasons for the timing of such a change will help us build better earthquake models. 

\section{Data and Resources}

We used the Advanced National Seismic System (ANSS) earthquake catalog made publicly available by the Northern California Earthquake Data Center at \url{www.ncedc.org} in the period from 1 January 1981 until 1 April 2009 with magnitude $ m \geq 2$ and in the spatial region defined by the CSEP collection region, defined in Table 2 by \citet{SchorlemmerGerstenberger2007}. For section \ref{sec:NBD}, we used the ANSS catalog in the CSEP testing region from 1 January 1932 until 1 April 2009. 

\section{Acknowledgments}

We thank A. Christophersen and J. Woessner for stimulating discussions. MJW is supported by the EXTREMES project of ETH's Competence Center Environment and Sustainability (CCES). AH is supported by the European Commission under grant TRIGS-043251, and by the French National Research Agency under grant  ASEISMIC. DDJ and YYK appreciate support from the National Science Foundation through grant EAR-0711515, as well as from the Southern California Earthquake Center (SCEC). SCEC is funded by NSF Cooperative Agreement EAR-0529922 and the U.S. Geological Survey (USGS) Cooperative Agreement 07HQAG0008. MJW thanks SCEC for travel support. Publication 0000, SCEC.


\begin{thebibliography}{53}
\providecommand{\natexlab}[1]{#1}
\expandafter\ifx\csname urlstyle\endcsname\relax
  \providecommand{\doi}[1]{doi:\discretionary{}{}{}#1}\else
  \providecommand{\doi}{doi:\discretionary{}{}{}\begingroup
  \urlstyle{rm}\Url}\fi

\bibitem[{\textit{Aki}(1965)}]{Aki1965}
Aki, K. (1965), Maximum likelihood estimate of b in the formula log {N}=a-b{M}
  and its confidence limits, \textit{Bull. Earthq. Res. Inst., Univ. Tokyo},
  \textit{43}, 237--239.

\bibitem[{\textit{Bird and Kagan}(2004)}]{BirdKagan2004}
Bird, P., and Y.~Y. Kagan (2004), Plate-tectonic analysis of shallow
  seismicity: Apparent boundary width, beta, corner magnitude, coupled
  lithosphere thickness, and coupling in seven tectonic settings, \textit{Bull.
  Seismol. Soc. Am.}, \textit{94}(6), 2380--2399, plus electronic supplement.

\bibitem[{\textit{Christophersen and Smith}(2008)}]{ChristophersenSmith2008}
Christophersen, A., and E.~G.~C. Smith (2008), {Foreshock Rates from Aftershock
  Abundance}, \textit{Bull. Seismol. Soc. Am.}, \textit{98}(5), 2133--2148,
  \doi{10.1785/0120060143}.

\bibitem[{\textit{Console et~al.}(2003)\textit{Console, Murru, and
  Lombardi}}]{Console-et-al2003b}
Console, R., M.~Murru, and A.~M. Lombardi (2003), Refining earthquake
  clustering models, \textit{J. Geophys. Res.}, \textit{108}(B10), 2468,
  \doi{10.1029/2002JB002123}.

\bibitem[{\textit{{Daley} and Vere-Jones}(2004)}]{DaleyVereJones2004}
{Daley}, D.~J., and D.~Vere-Jones (2004), Scoring probability forecasts for
  point processes: the entropy score and information gain, \textit{J. Appl.
  Prob.}, \textit{41A}, 297--312.

\bibitem[{\textit{Felzer et~al.}(2002)\textit{Felzer, Becker, Abercrombie,
  Ekstrom, and Rice}}]{Felzer-et-al2002}
Felzer, K.~R., T.~W. Becker, R.~E. Abercrombie, G.~Ekstrom, and J.~R. Rice
  (2002), Triggering of the 1999 {M}w 7.1 {H}ector {M}ine earthquake by
  aftershocks of the 1992 {M}w 7.3 {L}anders earthquake, \textit{J. Geophys.
  Res.}, \textit{107}(B09), \doi{10.1029/2001JB000911}.

\bibitem[{\textit{Felzer et~al.}(2004)\textit{Felzer, Abercrombie, and
  Ekstrom}}]{Felzer-et-al2004}
Felzer, K.~R., R.~E. Abercrombie, and G.~Ekstrom (2004), {A Common Origin for
  Aftershocks, Foreshocks, and Multiplets}, \textit{Bull. Seismol. Soc. Am.},
  \textit{94}(1), 88--98, \doi{10.1785/0120030069}.

\bibitem[{\textit{Field}(2007)}]{Field2007}
Field, E.~H. (2007), {A Summary of Previous Working Groups on California
  Earthquake Probabilities}, \textit{Bull. Seismol. Soc. Am.}, \textit{97}(4),
  1033--1053, \doi{10.1785/0120060048}.

\bibitem[{\textit{Gutenberg and Richter}(1944)}]{GutenbergRichter1944}
Gutenberg, B., and C.~F. Richter (1944), Frequency of earthquakes in
  {C}alifornia, \textit{Bull. Seis. Soc. Am.}, \textit{34}, 184--188.

\bibitem[{\textit{Hainzl and Ogata}(2005)}]{HainzlOgata2005}
Hainzl, S., and Y.~Ogata (2005), {Detecting fluid signals in seismicity data
  through statistical earthquake modeling}, \textit{J. Geophys. Res.},
  \textit{110, B05S07}, \doi{10.1029/2004JB003247}.

\bibitem[{\textit{Hainzl et~al.}(2008)\textit{Hainzl, Christophersen, and
  Enescu}}]{Hainzl-et-al2008}
Hainzl, S., A.~Christophersen, and B.~Enescu (2008), {Impact of Earthquake
  Rupture Extensions on Parameter Estimations of Point-Process Models},
  \textit{Bull. Seismol. Soc. Am.}, \textit{98}(4), 2066--2072,
  \doi{10.1785/0120070256}.

\bibitem[{\textit{Helmstetter}(2003)}]{Helmstetter2003}
Helmstetter, A. (2003), Is earthquake triggering driven by small earthquakes?,
  \textit{Phys. Rev. Lett.}, \textit{91}(5), 058,501,
  \doi{10.1103/PhysRevLett.91.058501}.

\bibitem[{\textit{Helmstetter and Sornette}(2002)}]{HelmstetterSornette2002}
Helmstetter, A., and D.~Sornette (2002), Subcritical and supercritical regimes
  in epidemic models of earthquake aftershocks, \textit{J. Geophys. Res.},
  \textit{107}(B10), 2237, \doi{10.1029/2001JB001580}.

\bibitem[{\textit{Helmstetter and Sornette}(2003)}]{HelmstetterSornette2003d}
Helmstetter, A., and D.~Sornette (2003), Predictability in the {ETAS} model of
  interacting triggered seismicity, \textit{J. Geophys. Res.},
  \textit{108}(B10), 2482, \doi{10.1029/2003JB002485}.

\bibitem[{\textit{Helmstetter et~al.}(2005)\textit{Helmstetter, Kagan, and
  Jackson}}]{Helmstetter-et-al2005}
Helmstetter, A., Y.~Y. Kagan, and D.~D. Jackson (2005), Importance of small
  earthquakes for stress transfers and earthquake triggering, \textit{J.
  Geophys. Res.}, \textit{110}, \doi{10.1029/2004JB003286}.

\bibitem[{\textit{Helmstetter et~al.}(2006)\textit{Helmstetter, Kagan, and
  Jackson}}]{Helmstetter-et-al2006}
Helmstetter, A., Y.~Y. Kagan, and D.~D. Jackson (2006), Comparison of
  short-term and time-independent earthquake forecast models for southern
  california, \textit{Bull. Seismol. Soc. Am.}, \textit{96}(1),
  \doi{10.1785/0120050067}.

\bibitem[{\textit{Helmstetter et~al.}(2007)\textit{Helmstetter, Kagan, and
  Jackson}}]{Helmstetter-et-al2007}
Helmstetter, A., Y.~Y. Kagan, and D.~D. Jackson (2007), High-resolution
  time-independent grid-based forecast for {M} ${\>}= 5$ earthquakes in
  {California}, \textit{Seismol. Res. Lett.}, \textit{78}(1), 78--86,
  \doi{10.1785/gssrl.78.1.78}.

\bibitem[{\textit{Jackson}(1996)}]{Jackson1996}
Jackson, D.~D. (1996), Hypothesis testing and earthquake prediction,
  \textit{Proc. Natl. Acad. Sci. USA}, \textit{93}, 3772--3775.

\bibitem[{\textit{Jackson and Kagan}(1999)}]{JacksonKagan1999}
Jackson, D.~D., and Y.~Y. Kagan (1999), Testable earthquake forecasts for 1999,
  \textit{Seismol. Res. Lett.}, \textit{70}(4), 393--403.

\bibitem[{\textit{Jordan}(2006)}]{Jordan2006}
Jordan, T.~H. (2006), Earthquake predictability: Brick by brick,
  \textit{Seismol. Res. Lett.}, \textit{77}(1), 3--6.

\bibitem[{\textit{Kagan}(1973)}]{Kagan1973}
Kagan, Y.~Y. (1973), Statistical methods in the study of seismic processes,
  \textit{Bull. Int. Stat. Inst.}, \textit{45(3)}, 437--453, with discussion.

\bibitem[{\textit{Kagan}(1991)}]{Kagan1991a}
Kagan, Y.~Y. (1991), Likelihood analysis of earthquake catalogs,
  \textit{Geophys. J. Intern.}, \textit{106}, 135--148.

\bibitem[{\textit{Kagan}(2004)}]{Kagan2004}
Kagan, Y.~Y. (2004), Short-term properties of earthquake catalogs and models of
  earthquake source, \textit{Bull. Seismol. Soc. Am.}, \textit{94}(4),
  1207--1228.

\bibitem[{\textit{Kagan and Jackson}(1994)}]{KaganJackson1994}
Kagan, Y.~Y., and D.~D. Jackson (1994), Long-term probabilistic forecasting of
  earthquakes, \textit{J. Geophys. Res.}, \textit{99}(B7), 13,685--13,700.

\bibitem[{\textit{Kagan and Knopoff}(1987)}]{KaganKnopoff1987}
Kagan, Y.~Y., and L.~Knopoff (1987), Statistical short-term earthquake
  prediction, \textit{Science}, \textit{236}(4808), 1563--1567.

\bibitem[{\textit{Lennartz et~al.}(2008)\textit{Lennartz, Bunde, and
  Turcotte}}]{Lennartz-et-al2008}
Lennartz, S., A.~Bunde, and D.~L. Turcotte (2008), Missing data in aftershock
  sequences: Explaining the deviations from scaling laws, \textit{Physical
  Review E (Statistical, Nonlinear, and Soft Matter Physics)}, \textit{78}(4),
  041115, \doi{10.1103/PhysRevE.78.041115}.

\bibitem[{\textit{Marsan and Lengline}(2008)}]{MarsanLengline2008}
Marsan, D., and O.~Lengline (2008), {Extending Earthquakes' Reach Through
  Cascading}, \textit{Science}, \textit{319}(5866), 1076--1079,
  \doi{10.1126/science.1148783}.

\bibitem[{\textit{Ogata}(1988)}]{Ogata1988}
Ogata, Y. (1988), Statistical models for earthquake occurrence and residual
  analysis for point processes, \textit{J. Am. Stat. Assoc.}, \textit{83},
  9--27.

\bibitem[{\textit{Ogata}(1998)}]{Ogata1998}
Ogata, Y. (1998), Space-time point-process models for earthquake occurrences,
  \textit{Ann. Inst. Stat. Math.}, \textit{5}(2), 379--402.

\bibitem[{\textit{Ogata}(2004)}]{Ogata2004}
Ogata, Y. (2004), Space-time model for regional seismicity and detection of
  crustal stress changes, \textit{J. Geophys. Res.}, \textit{109}(B03308),
  \doi{10.1029/2003JB002621}.

\bibitem[{\textit{Peng et~al.}(2007)\textit{Peng, Vidale, Ishii, and
  Helmstetter}}]{Peng-et-al2007}
Peng, Z., J.~E. Vidale, M.~Ishii, and A.~Helmstetter (2007), Seismicity rate
  immediately before and after main shock rupture from high-frequency waveforms
  in {J}apan, \textit{J. Geophys. Res.}, \textit{112}(B03306),
  \doi{10.1029/2006JB004386}.

\bibitem[{\textit{Press et~al.}(1992)\textit{Press, Teukolsky, Vetterling, and
  Flannery}}]{Press-et-al1992}
Press, W.~H., S.~A. Teukolsky, W.~T. Vetterling, and B.~P. Flannery (1992),
  \textit{Numerical Recipes in Fortran: The Art of Scientific Computing}, 2nd.
  ed., 992 pp., Cambridge Univ. Press, New York, USA.

\bibitem[{\textit{Reasenberg}(1985)}]{Reasenberg1985}
Reasenberg, P. (1985), Second-order moment of central {C}alifornia seismicity,
  1969-82, \textit{J. Geophys. Res.}, \textit{90}, 5479--5495.

\bibitem[{\textit{Saichev and Sornette}(2005)}]{SaichevSornette2005b}
Saichev, A., and D.~Sornette (2005), Vere-{J}ones' self-similar branching
  model, \textit{Phys. Rev. E}, \textit{72}, 056,122.

\bibitem[{\textit{Saichev and
  Sornette}(2006{\natexlab{a}})}]{SaichevSornette2006}
Saichev, A., and D.~Sornette (2006{\natexlab{a}}), Renormalization of branching
  models of triggered seismicity from total to observable seismicity,
  \textit{Eur. Phys. J. B}, \textit{51}, 443--459.

\bibitem[{\textit{Saichev and
  Sornette}(2006{\natexlab{b}})}]{SaichevSornette2006c}
Saichev, A., and D.~Sornette (2006{\natexlab{b}}), Power law distribution of
  seismic rates: theory and data analysis, \textit{Eur. Phys. J. B},
  \textit{49}, 377--401, \doi{10.1140/epjb/e2006-00075-3}.

\bibitem[{\textit{Schorlemmer and
  Gerstenberger}(2007)}]{SchorlemmerGerstenberger2007}
Schorlemmer, D., and M.~Gerstenberger (2007), {RELM} testing center,
  \textit{Seismol. Res. Lett.}, \textit{78}(1), 30.

\bibitem[{\textit{Schorlemmer and Woessner}(2008)}]{SchorlemmerWoessner2008}
Schorlemmer, D., and J.~Woessner (2008), {Probability of Detecting an
  Earthquake}, \textit{Bull. Seismol. Soc. Am.}, \textit{98}(5), 2103--2117,
  \doi{10.1785/0120070105}.

\bibitem[{\textit{Schorlemmer et~al.}(2007)\textit{Schorlemmer, Gerstenberger,
  Wiemer, Jackson, and Rhoades}}]{Schorlemmer-et-al2007}
Schorlemmer, D., M.~C. Gerstenberger, S.~Wiemer, D.~D. Jackson, and D.~A.
  Rhoades (2007), Earthquake likelihood model testing, \textit{Seismol. Res.
  Lett.}, \textit{78}(1), 17.

\bibitem[{\textit{Schorlemmer et~al.}(2009)\textit{Schorlemmer, Zechar, Werner,
  Field, Jackson, Jordan, and {the RELM Working
  Group}}}]{Schorlemmer-et-al2009}
Schorlemmer, D., J.~D. Zechar, M.~J. Werner, E.~Field, D.~D. Jackson, T.~H.
  Jordan, and {the RELM Working Group} (2009), First results of the regional
  earthquake likelihood models experiment, \textit{Pure and Appl. Geophys., in
  press}.

\bibitem[{\textit{Sornette and
  Werner}(2005{\natexlab{a}})}]{SornetteWerner2005a}
Sornette, D., and M.~J. Werner (2005{\natexlab{a}}), Constraints on the size of
  the smallest triggering earthquake from the epidemic-type aftershock sequence
  model, {B}{\aa}th's law, and observed aftershock sequences, \textit{J.
  Geophys. Res.}, \textit{110}(B08304), \doi{10.1029/2004JB003535}.

\bibitem[{\textit{Sornette and
  Werner}(2005{\natexlab{b}})}]{SornetteWerner2005b}
Sornette, D., and M.~J. Werner (2005{\natexlab{b}}), Apparent clustering and
  apparent background earthquakes biased by undetected seismicity, \textit{J.
  Geophys. Res.}, \textit{110}(B09303), \doi{10.1029/2005JB003621}.

\bibitem[{\textit{Vere-Jones}(1970)}]{VereJones1970}
Vere-Jones, D. (1970), Stochastic models for earthquake occurrence, \textit{J.
  Roy. Stat. Soc. Series B (Methodological)}, \textit{32}(1), 1--62 (with
  discussion).

\bibitem[{\textit{Wells and Coppersmith}(1994)}]{WellsCoppersmith1994}
Wells, D.~L., and K.~J. Coppersmith (1994), {New empirical relationships among
  magnitude, rupture length, rupture width, rupture area, and surface
  displacement}, \textit{Bull. Seismol. Soc. Am.}, \textit{84}(4), 974--1002.

\bibitem[{\textit{Werner}(2007)}]{Werner2007}
Werner, M.~J. (2007), On the fluctuations of seismicity and uncertainties in
  earthquake catalogs: Implications and methods for hypothesis testing, Ph.D.
  thesis, Univ. California, Los Angeles.

\bibitem[{\textit{Werner and Sornette}(2008)}]{WernerSornette2008a}
Werner, M.~J., and D.~Sornette (2008), Magnitude uncertainties impact seismic
  rate estimates, forecasts and predictability experiments, \textit{J. Geophys.
  Res.}, \doi{10.1029/2007JB005427}.

\bibitem[{\textit{{Wiemer}}(2000)}]{Wiemer2000}
{Wiemer}, S. (2000), {Introducing probabilistic aftershock hazard mapping},
  \textit{Geophys. Res. Lett.}, \textit{27}, 3405--3408,
  \doi{10.1029/2000GL011479}.

\bibitem[{\textit{{Wiemer} and {Katsumata}}(1999)}]{WiemerKatsumata1999}
{Wiemer}, S., and K.~{Katsumata} (1999), {Spatial variability of seismicity
  parameters in aftershock zones}, \textit{J. Geophys. Res.}, \textit{104},
  13,135--13,152, \doi{10.1029/1999JB900032}.

\bibitem[{\textit{Zechar et~al.}(2009{\natexlab{a}})\textit{Zechar,
  Schorlemmer, Liukis, Yu, Euchner, Maechling, and Jordan}}]{Zechar-et-al2009}
Zechar, J.~D., D.~Schorlemmer, M.~Liukis, J.~Yu, F.~Euchner, P.~J. Maechling,
  and T.~H. Jordan (2009{\natexlab{a}}), The {C}ollaboratory for the {S}tudy of
  {E}arthquake {P}redictability perspective on computational earthquake
  science, \textit{Concurrency and Computation: Practice and Experience},
  \doi{10.1002/cpe.1519}.

\bibitem[{\textit{Zechar et~al.}(2009{\natexlab{b}})\textit{Zechar,
  Gerstenberger, and Rhoades}}]{Zechar-et-al2009b}
Zechar, J.~D., M.~C. Gerstenberger, and D.~Rhoades (2009{\natexlab{b}}),
  Likelihood-based tests for evaluating the spatial and magnitude component of
  earthquake rate forecasts, \textit{Bull. Seismol. Soc. Am., in revision}.

\bibitem[{\textit{Zhuang et~al.}(2002)\textit{Zhuang, Ogata, and
  Vere-Jones}}]{Zhuang-et-al2002}
Zhuang, J., Y.~Ogata, and D.~Vere-Jones (2002), Stochastic declustering of
  space-time earthquake occurrences, \textit{J. Am. Stat. Assoc.}, \textit{97},
  369--380.

\bibitem[{\textit{Zhuang et~al.}(2004)\textit{Zhuang, Ogata, and
  Vere-Jones}}]{Zhuang-et-al2004}
Zhuang, J., Y.~Ogata, and D.~Vere-Jones (2004), Analyzing earthquake clustering
  features by using stochastic reconstruction, \textit{J. Geophys. Res.},
  \textit{109}(B05301), \doi{10.1029/2003JB002879}.

\bibitem[{\textit{{Zhuang} et~al.}(2008)\textit{{Zhuang}, {Christophersen},
  {Savage}, {Vere-Jones}, {Ogata}, and {Jackson}}}]{Zhuang-et-al2008}
{Zhuang}, J., A.~{Christophersen}, M.~K. {Savage}, D.~{Vere-Jones}, Y.~{Ogata},
  and D.~D. {Jackson} (2008), {Differences between spontaneous and triggered
  earthquakes: Their influences on foreshock probabilities}, \textit{J.
  Geophys. Res.}, \textit{113}(B12), \doi{10.1029/2008JB005579}.

\end{thebibliography}

\begin{centering}
\section*{Author's Affiliations, Addresses}
\end{centering}

\noindent  Maximilian J. Werner, Swiss Seismological Service, Institute of Geophysics, ETH Zurich, Sonneggstr. 5, 8092 Zurich, Switzerland. 

\noindent Agn\`{e}s Helmstetter, Laboratoire de G\'eophysique Interne et Tectonophysique, Universit\'e Joseph Fourier and Centre National de la Recherche Scientifique, BP 53, 38041 Grenoble, France.

\noindent David D. Jackson, Department of Earth and Space Sciences, University of California, Los Angeles, CA 90095-1567.

\noindent Yan Y. Kagan, Department of Earth and Space Sciences, University of California, Los Angeles, CA 90095-1567.

\clearpage


\begin{centering}
\section*{Tables}
\end{centering}

\begin{table}[htbp]
\caption{ {\bf Results of the optimization of the spatial background model:}  \newline
Gaussian (gs) (\ref{eq:Kgauss}) or power-law (pl) kernel (\ref{eq:Kpl}). Input catalog: declustered ANSS catalog. Target catalog: ANSS catalog. $N$: number of earthquakes. $L$: log-likelihood  (\ref{eq:L}), $G$: probability gain per earthquake over a spatially uniform model  (\ref{eq:G}). $n_v$: optimal number of neighbors in the bandwidth of the smoothing kernel. The superscript $^!$ denotes that $n_v$ was constrained, and for $*$, the target $m_{min}=3$. }
  \centering
  \begin{tabular}{@{} cc|cccc|ccc|ccc @{}}
    \hline
    No. & K & \multicolumn{4}{c|}{Input Catalog} & \multicolumn{3}{c|}{Target Catalog}  & \multicolumn{3}{c}{Results} \\
     & & $t_1$ & $t_2$ & $m_{min}$ & N & $t_1$ & $t_2$ &  N & L & G & $n_v$ \\ 
    \hline
     &  & \multicolumn{4}{c|}{} & \multicolumn{3}{c|}{$m_{min}=4.95$}  & \multicolumn{3}{c}{} \\
    \hline
    1 & gs & 1981 & 2003 & 2. & 81,455 & 2004 & 2008  & 25 & -131.2 & 5.13 & 2 \\ 
    2 & pl & 1981 & 2003 & 2. & 81,455 &2004 & 2008  & 25 & -131.3 & 5.11 & 3 \\ 
    3 & gs & 1981 & 2003 & 2.5 & 25,407 & 2004 & 2008  & 25 & -127.8 & 5.88 & 1 \\ 
    4 & pl & 1981 & 2003 & 2.5 & 25,407 & 2004 & 2008  & 25 & -128.4 & 5.73 & 1 \\ 
    5 & gs & 1981 & 2003 & 3. & 8,090 & 2004 & 2008  & 25 & -124.7 & 6.73 & 2 \\ 
    6 & pl & 1981 & 2003 & 3. & 8,090 & 2004 & 2008  & 25 & -125.3 & 6.49 & 2 \\ 
    7 & gs & 1981 & 2003 & 3.5 & 2,383 & 2004 & 2008  & 25 & -125.8 & 6.37 & 3 \\ 
    8 & pl & 1981 & 2003 & 3.5 & 2,383 & 2004 & 2008  & 25 & -127.1 & 6.04 & 1 \\ 
    9 & gs & 1981 & 2003 & 4. & 813 & 2004 & 2008  & 25 &  -125.1 & 6.53 &  3 \\ 
    10 & pl & 1981 & 2003 & 4. & 813 & 2004 & 2008  & 25 &  -126.6 & 6.17 &  2 \\ 
    11 & gs & 1981 & 2003 & 4.5 & 259 & 2004 & 2008  & 25 & -130.3 & 5.31 & 1 \\ 
    12 & pl & 1981 & 2003 & 4.5 & 259 & 2004 & 2008  & 25 & -131.3 & 5.11 & 1 \\ 
    13 & gs & 1981 & 2003 & 5. & 101 & 2004 & 2008  & 25 & -150.6 & 2.36 & 1 \\ 
    14 & pl & 1981 & 2003 & 5. & 101 & 2004 & 2008  & 25 & -148.1 & 2.61 & 1 \\ 
    15 &  gs & 1981 & 2003 & 5.5 & 38 & 2004 & 2008  & 25 & -161.7 & 1.51 & 3 \\ 
    16 &  pl & 1981 & 2003 & 5.5 & 38 & 2004 & 2008  & 25 & -154.6 & 2.01 & 1 \\ 
    17 & pl & 1981 & 1998 & 2. & 65,962 & 1999 & 2003  & 22 &  -131.2 &  2.68 &  1 \\ 
    18 & pl & 1981 & 1993 & 2. & 49,063 & 1994 & 1998  & 32 &  -191.5 & 2.36  &   2 \\ 
    19 & pl & 1981 & 1988 & 2. & 31,569 & 1989 & 1993  & 52 &  -271.0 &  2.86 &  9 \\ 
    20 & pl & 1981 & 4/2009 & 2. & 96,186 & 2004 & 4/2009  & 25 & -123.8 & 6.90 & 2 \\ 
    $21^{!}$ & pl & 1981 & 4/2009 & 2. & 96,186 & 2004 & 4/2009  & 25& -124.4 & 6.73 & $6^{!}$ \\ 
    \hline
     &  & \multicolumn{4}{c|}{} & \multicolumn{3}{c|}{$m_{min}=3$}  & \multicolumn{3}{c}{} \\
    \hline
    22* & pl & 1981 & 8/23/2005 & 2. & 86,409 & 1996 & 8/23/2005  & 2763 & -3215.5 & 7.08 & 1 \\ 
    23$^{!}$* & pl & 1981 & 8/23/2005 & 2. & 86,409 & 1996 & 8/23/2005  & 2763 & -3309.5 & 6.85 & $6^{!}$ \\ 
    \hline
   \end{tabular}
  
  \label{tab:SmoOpt}
\end{table}

\clearpage

\begin{table}[htbp]
\caption{ {\bf Parameter estimates of the ETAS model}: \newline 
Input catalog: ANSS catalog 1/1/1981 to 4/1/2009 in the CSEP collection region (167,528 earthquakes $m\geq2$). Target catalog: ANSS catalog 1/1/1986 to 4/1/2009 in the CSEP testing region (1,521 earthquakes for $m_{min}=3.95$). Spatial background model 21 from Table \ref{tab:SmoOpt}. Reference model: time-independent forecast model 21 and the average number $\mu_{TI}=0.18$ of daily $m>3.95$. $^{\ddagger}$: Using the corrective term $\rho^{\star}(m)$ of equation (\ref{eq:rhostar}). $^{\S (m=M)}$: Early-aftershock-smoothing kernel for quakes above $m \geq M$ [default $m=5.5$]. $^{\S (T=t)}$: Early-aftershock-smoothing kernel for aftershocks occurring up to $t$ days after a large event [default $T=2$ days]. }
\begin{center}
  \begin{tabular}{@{} ccccccccccccc @{}}
    \hline
    Model & kr & $m_{min}$ & $\alpha$ & $p$ & $k$ & $\mu_s$ & $f_d$ & $L_{ETAS}$ &$L_{TI}$ & $N_{obs}$ & $N_{pred}$ & $G$ \\ 
    \hline
    $1$ 			& gs & 3.95 & 0.84 & 1.28 & 0.34 & 0.083 & 0.89 & -15,977 &  -18,749 & 1,521 & 1,529.0 & 6.19 \\  
    $2$ 			& pl & 3.95 & 0.82 & 1.26 & 0.39 & 0.072 & 0.08 & -15,944 & -18,729 & 1,521 & 1,528.8 & 6.32 \\  
    $3^{\ddagger}$ 	& gs & 3.95 & 0.81 & 1.30 & 0.33 & 0.084 & 0.78 & -15,991 &  -18,729 & 1,521 & 1,518.9 &  6.13 \\  
    $4^{\ddagger}$ 	& pl & 3.95 & 0.81 & 1.27 & 0.32 & 0.075 & 0.17 & -15,957 &  -18,729 & 1,521 & 1,483.5 &  6.27 \\  
    $5^{\S (m=5)}$ 	& gs & 3.95 & 0.84  & 1.27 & 0.33 & 0.081 & 0.94 & -15,976 &  -18,729 & 1,521 & 1,511.3 & 6.19 \\  
    $6^{\S (m=5)}$ 	& pl & 3.95 & 0.82  & 1.27 & 0.42 & 0.074 & 0.09 & -15,945 &  -18,729 & 1,521 & 1,533.6 & 6.31 \\  
    $7^{\S (m=6)}$ 	& gs & 3.95 & 0.84 & 1.27 & 0.35 & 0.081 & 0.59 & -15,972 &  -18,729 & 1,521 & 1,529.8 & 6.20 \\  
    $8^{\S (m=6)}$ 	& pl & 3.95 & 0.80 & 1.28 & 0.46 & 0.073 & 0.16 & -15,948 &  -18,729 & 1,521 & 1,518.0 & 6.31 \\  
    $9^{\S (m=\infty)}$ & gs & 3.95 & 0.77 & 1.28 & 0.49 & 0.078 & 0.51 & -16,013 &  -18,729 & 1,521 & 1,507.8 & 6.04 \\  
    $10^{\S (m=\infty)}$ & pl & 3.95 & 0.71 & 1.20 & 0.60 & 0.057 & 0.25 & -16,008 &  -18,729 & 1,521 & 1,586.9 & 6.06 \\  
    $11^{\S (T=1)}$ & gs & 3.95 & 0.83  & 1.28 & 0.36 & 0.079 & 0.74 & -15,980 & -18,729  & 1,521 & 1,496.9 & 6.17 \\  
    $12^{\S (T=1)}$ & pl & 3.95 & 0.82  & 1.27 & 0.41 & 0.074 & 0.11 & -15,945 & -18,729  & 1,521 & 1,522.1 & 6.32 \\  
    $13^{\S (T=3)}$ & gs & 3.95 &  0.84 & 1.27 & 0.34 & 0.081 & 0.75 & -15,975 & -18,729  & 1,521 & 1,520.7 & 6.19 \\  
    $14^{\S (T=3)}$ & pl & 3.95 &  0.81 & 1.28 & 0.42 & 0.074 & 0.10 & -15,943 & -18,729  & 1,521 & 1,513.6 & 6.34 \\  
     \hline
  \end{tabular}
\label{tab:optETASm4}
\end{center}
\end{table}

\clearpage

\section{E-Supplements}

\subsection{Supplement 1: Removal of Explosions from the ANSS Catalog}

\begin{figure}[h]
\centering
\includegraphics[draft=\IsDraft,width=\halfwidth,keepaspectratio=true,clip]{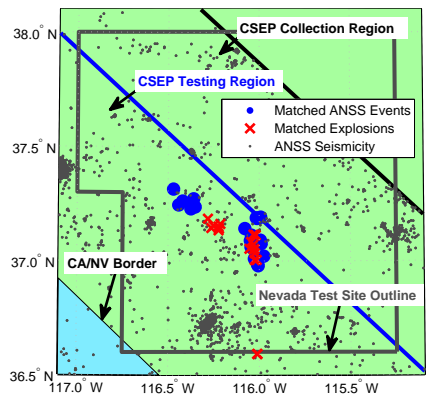}
\caption{\label{fig:map} We matched events in the ANSS catalog from 1981 until 2009 with known underground nuclear explosions: Matched ANSS events (blue filled circles), matched explosions (red crosses) and unmatched ANSS earthquakes (grey points). Also shown are an approximate outline of the Nevada nuclear test site (grey) in relation to the CSEP testing (blue line) and collection regions (black line). 
}
\end{figure}

\begin{figure}[ht]
\centering
\includegraphics[draft=\IsDraft,width=\halfwidth,keepaspectratio=true,clip]{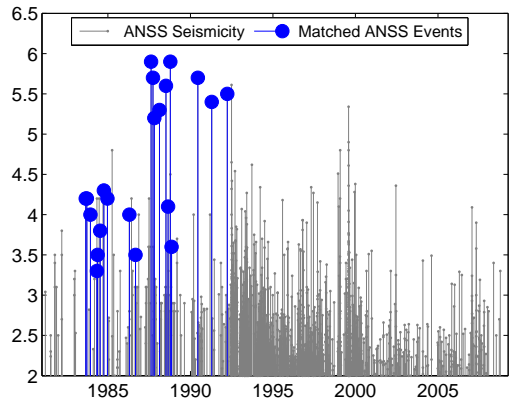}
\caption{\label{fig:Exps} Identified explosions (large blue filled circles) and earthquakes (small grey dots) in the approximate Nevada Test Site area in the ANSS catalog from 1981 until 2009 (see Figure \ref{fig:map}). 
}
\end{figure}

The Sandia National Lab published a list of official underground nuclear explosions at \url{nuclearweaponarchive.org/usa/tests/nevada.html}. We matched explosions to entries in the earthquake catalog whenever the events occurred within 12 seconds and within $\pm 0.5$ degree latitude and longitude. Increasing the time criterion up to 18 minutes or the space criterion up to $\pm 1.5$ does not change the matched events. On the other hand, 19 events are matched by a spatial constraint of $0.2$ degrees. The two extra matches come from two explosions that occur on the southern edge of the Nevada Test Site (see Figure \ref{fig:map}). These are well-matched in time by large and shallow events that are located at the center of the Site in the usual region of underground explosions. It is possible that either the locations of these two explosions contain an error (as they are exactly co-located in latitude and farther South than all other explosions), or that they triggered shallow collapses of the domes of past explosions. Because we did not want to forecast explosions, nor events caused by the collapse of domes after the explosions, nor quakes triggered by explosions, we assumed a location uncertainty larger than usual. Nine of these events were large $m\geq 5$, none had identifiable aftershock sequences and all occurred between 1984 and 1992 (see Figure S\ref{fig:Exps}). We deleted these 21 events from the earthquake catalog.




\clearpage

\subsection{Supplement 2: Optimizing the Spatial Smoothing for Different Magnitude Thresholds of the Target Catalog}

To show that our smoothing method performs well for various magnitude thresholds of the target catalog, we present further results in Table \ref{tab:OptTarMags}. When we increase the target threshold from $m_{min}=2$ to $m_{min}=5$, the probability gain per quake remains roughly constant. We do observe a drop for target $m_{min}=5.5$, but there are only 2 events that determine this score. At the same time, when we increase the input magnitude threshold from $m_{min}=2$ to $m_{min}=5.5$, the probability gain decreases quickly, suggesting that small earthquakes can help forecast all seismicity, whether small events or large ones. Different target periods obtain different gains, but the fluctuations are much smaller than in the case of $m_{min}=4.95$ (see Table \ref{tab:SmoOpt}). One explanation for the smaller fluctuations is the increased number of target events, but this needs to be investigated further.  

\begin{table}[htbp]
\caption{ {\bf Results of the optimization of the spatial background model:}  \newline
Input catalog: Declustered ANSS catalog. Target catalog: ANSS catalog. $N$: number of earthquakes. $L$: log-likelihood, $G$: probability gain per earthquake over a spatially uniform model. $n_v$: optimal number of neighbors to include in the bandwidth of the smoothing kernel. The superscript $^!$ denotes that $n_v$ was constrained.
}
\centering
  \begin{tabular}{@{} cc|cccc|cccc|ccc @{}}
    \hline
    No. & K & \multicolumn{4}{c|}{Input Catalog} & \multicolumn{4}{c|}{Target Catalog}  & \multicolumn{3}{c}{Results} \\
     & & $t_1$ & $t_2$ & $m_{min}$ & N & $t_1$ & $t_2$ &  $m_{min}$ & N & L & G & $n_v$\\ 
    \hline
    1 & pl & 1981 & 2003 & 2. & 81,455 &2004 & 2008  & 2.	& 18,491 & -23,971.0 & 5.65 & 2 \\ 
    2 & pl & 1981 & 2003 & 2. & 81,455 &2004 & 2008  & 2.5	& 5784 & -8739.9 & 5.58 & 2 \\ 
    3 & pl & 1981 & 2003 & 2. & 81,455 &2004 & 2008  & 3.	&   1726 & -3630.3 & 5.03 & 2 \\ 
    4 & pl & 1981 & 2003 & 2. & 81,455 &2004 & 2008  & 3.5	&    571   & -1571.9 & 4.59 & 2 \\ 
    5 & pl & 1981 & 2003 & 2. & 81,455 &2004 & 2008  & 4.	&    198   & -664.7 & 5.39 & 3 \\ 
    6 & pl & 1981 & 2003 & 2. & 81,455 &2004 & 2008  & 4.5	&  66 & -298.0 & 4.50 & 3 \\ 
    7 & pl & 1981 & 2003 & 2. & 81,455 &2004 & 2008  & 5.	& 21 & -111.1 & 5.84 & 3 \\ 
    8 & pl & 1981 & 2003 & 2. & 81,455 &2004 & 2008  & 5.5	& 2 & -16.2 & 3.19 & 2 \\ 
    \hline
    3 & pl & 1981 & 2003 & 2. 	& 81,455 &2004 & 2008  & 3.	&   1726 & -3630.3 & 5.03 & 2 \\ 
    9 & pl & 1981 & 2003 & 2.5	& 25,407 & 2004 & 2008  & 3.	& 1726 & -3613.9 & 5.08 & 2 \\ 
    10 & pl & 1981 & 2003 & 3. 	& 8,090 & 2004 & 2008  	& 3.	& 1726 & -3712.7 & 4.79 & 2 \\ 
    11 & pl & 1981 & 2003 & 3.5	& 2,383 & 2004 & 2008  	& 3.	& 1726 & -3971.5 & 4.13 & 3 \\ 
    12 & pl & 1981 & 2003 & 4.	& 813 & 2004 & 2008  	& 3.	& 1726 &  -4129.7 & 3.77 &  2 \\ 
    13 & pl & 1981 & 2003 & 4.5	& 259 & 2004 & 2008  	& 3.	&1726 & -5036.9 & 2.23 & 1 \\ 
    14 & pl & 1981 & 2003 & 5.	& 101 & 2004 & 2008  	& 3.	& 1726 & -5648.8 & 1.56 & 2 \\ 
    15 & pl & 1981 & 2003 & 5.5	& 38 & 2004 & 2008  	& 3.	&1726 & -5951.6 & 1.31 & 1 \\ 
    \hline
    16 & pl & 1981 & 1998 & 2. 	& 65,962 & 1999 & 2003  	& 3.	& 2766 &  -6790.7 &  3.68 & 3  \\ 
    17 & pl & 1981 & 1993 & 2. 	& 49,063 & 1994 & 1998  	& 3.	& 2664 &  -5940.5 & 5.69  &   2 \\ 
    18 & pl & 1981 & 1988 & 2. 	& 31,569 & 1989 & 1993  	& 3.	& 4290 &  -9185.2 &  4.52 &  5 \\ 
    \hline
    19 & pl & 1981 & 4/2009 & 2.& 96,186 & 2004 & 4/2009  & 3.& 1814 & -2977.2 & 7.63 & 1 \\ 
    $20^!$ & pl & 1981 & 4/2009 & 2. & 96,186 & 2004 & 4/2009  & 3. & 1814& -30.29.9 & 7.41 & $6$ \\ 
 \hline
  \label{tab:OptTarMags}
   \end{tabular}
   \end{table}

\newpage

\subsection{Supplement 3: Long-term Forecast for the CSEP Mainshock-Only Group}

The RELM experiment designated two forecast groups: the mainshock-only group and the mainshock/aftershock group \citep{Schorlemmer-et-al2007, SchorlemmerGerstenberger2007}. The latter group comprises all earthquakes, while the former only includes so-called mainshocks selected by the Reasenberg declustering algorithm \citep{Reasenberg1985}. Because this declustering method is relatively arbitrary, it is difficult to argue that forecasts should be evaluated against these events only. Nevertheless, we submit to the CSEP competition an updated 5-year mainshock-only forecast to provide a simple null hypothesis for this particular group, too. This supplement describes how we calculated the forecast. 

We first preceded exactly as for the 5-year long term forecast of all earthquakes, but we optimized the spatial smoothing on a declustered target catalog. Table \ref{tab:OptTarMagsDecl} summarizes the log-likelihood scores and gains over a spatially uniform forecast. Comparing Table  \ref{tab:OptTarMagsDecl} to Table  \ref{tab:SmoOpt}, which lists the scores obtained on the original catalog, we find that the gains are uniformly a little lower, except for model 11, whose gain increased significantly from $2.68$ to $4.55$. All other models obtained slightly lower gains on the declustered target catalog than on the original target catalog, but the conclusion that large earthquakes occur on average in the locations of small ones remains valid. As before, our favorite spatial model is model 14 fixed $n_v=6$, since it uses all data and provides a slightly smoother model than one would obtain from the mean optimal smoothing parameter over various target periods. 

\begin{table}[htbp]
\caption{ {\bf Results of the optimization of the spatial background model for the mainshock-only group:}  \newline
Input catalog: declustered ANSS catalog. Target catalog: declustered ANSS catalog. $N$: number of earthquakes. $L$: log-likelihood, $G$: probability gain per earthquake over a spatially uniform model. $n_v$: optimal number of neighbors to include in the bandwidth of the smoothing kernel. The superscript $^!$ denotes that $n_v$ was constrained.
}
  \centering
  \begin{tabular}{@{} cc|cccc|ccc|ccc @{}}
    \hline
    No. & K & \multicolumn{4}{c|}{Input Catalog} & \multicolumn{3}{c|}{Target Catalog}  & \multicolumn{3}{c}{Results} \\
     & & $t_1$ & $t_2$ & $M_{min}$ & N & $t_1$ & $t_2$ &  N & L & G & $n_v$ \\ 
    \hline
     &  & \multicolumn{4}{c|}{} & \multicolumn{3}{c|}{$M_{min}=4.95$}  & \multicolumn{3}{c}{} \\
    \hline
    1$^{decl}$ & gs & 1981 & 2003 & 2. & 81,455 & 2004 & 2008  & 20 & -110.3 & 4.60 & 2 \\ 
    2$^{decl}$ & {pl} & 1981 & 2003 & 2. & 81,455 &2004 & 2008  & 20 & -110.4 & 4.57 & 3 \\ 
    3$^{decl}$ & pl & 1981 & 2003 & 2.5 & 25,407 & 2004 & 2008  & 20 &  -108.6 & 5.01 & 1 \\ 
    4$^{decl}$ & pl & 1981 & 2003 & 3. & 8,090 & 2004 & 2008  & 20 & -105.9 & 5.73 & 2 \\ 
    5$^{decl}$ & pl & 1981 & 2003 & 3.5 & 2,383 & 2004 & 2008  & 20 & -106.5 & 5.57 & 1 \\ 
    6$^{decl}$ & pl & 1981 & 2003 & 4. & 813 & 2004 & 2008  & 20 &  -107.9 & 5.19 &  2 \\ 
    7$^{decl}$ & pl & 1981 & 2003 & 4.5 & 259 & 2004 & 2008  & 20 & -111.5 & 4.33 & 1 \\ 
    8$^{decl}$ & pl & 1981 & 2003 & 5. & 101 & 2004 & 2008  & 20 & -123.0 & 2.44 & 1 \\ 
    9$^{decl}$ &  pl & 1981 & 2003 & 5.5 & 38 & 2004 & 2008  & 20 & -127.1 & 1.99 & 1 \\ 
    10$^{decl}$ & pl & 1981 & 1998 & 2. & 65,962 & 1999 & 2003  & 15 & -85.8 & 4.55 & 6 \\ 
    11$^{decl}$ & pl & 1981 & 1993 & 2. & 49,063 & 1994 & 1998  & 20 &  -122.2 & 2.66  &   2 \\ 
    12$^{decl}$ & pl & 1981 & 1988 & 2. & 31,569 & 1989 & 1993  & 18 &  -123.7 &  1.33 &  9 \\ 
    13$^{decl}$ & pl & 1981 & 4/2009 & 2. & 96,186 & 2004 & 4/2009  & 20 & -104.0 & 6.30 & 2 \\ 
    $14^{decl, !}$ & pl & 1981 & 4/2009 & 2. & 96,186 & 2004 & 4/2009  & 20& -104.7 & 6.09 & $6^{!}$ \\ 
 \hline
   \end{tabular}
  \label{tab:OptTarMagsDecl}
\end{table}

\begin{figure}[ht]
\centering
\includegraphics[draft=\IsDraft,width=\halfwidth,keepaspectratio=true,clip]{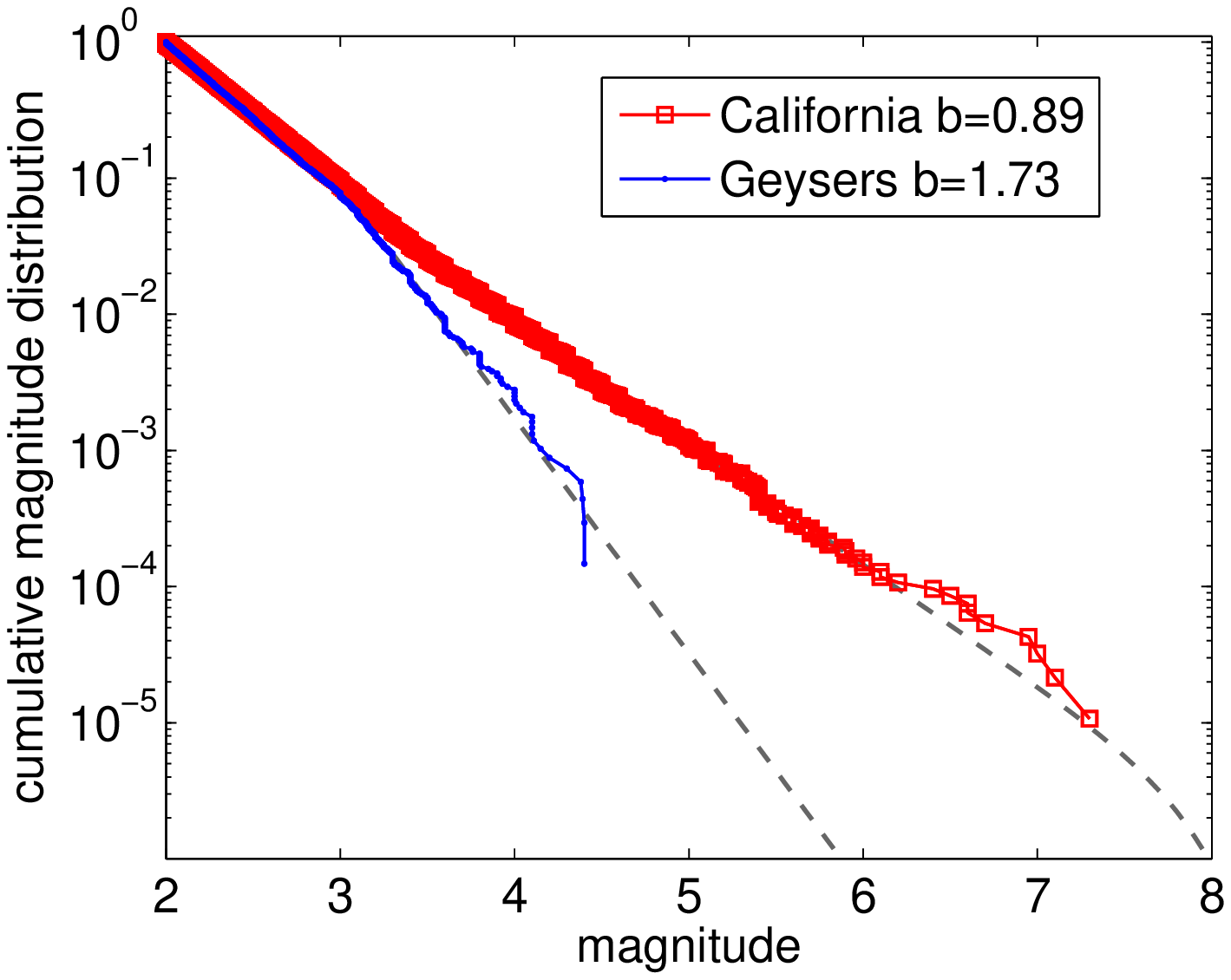}
\caption{\label{fig:MDe} Magnitude distribution of (declustered) earthquakes in the Geysers region (blue) and the remainder of California (red squares) from 1 January 1981 to 1 April 2009. 
}
\end{figure}

Having obtained our normalized, spatial model, we used the tapered Gutenberg-Richter distribution (\ref{tGR}) with corner magnitude $8.0$ for the entire region. Based on the declustered target catalog, we estimated $b\approx0.89$ for California and $b\approx1.73$ for the Geysers region (see Figure \ref{fig:MDe} and section \ref{sec:MD}). To obtain the number of expected events, we counted the number of events in the declustered catalog and divided by its length. We expect a total of $N_{pred}=20.51$ mainshock events in the next five years from 2010 until 2014. 

\begin{figure}[ht]
\centering
\includegraphics[draft=\IsDraft,width=\mediumwidth,keepaspectratio=true,clip]{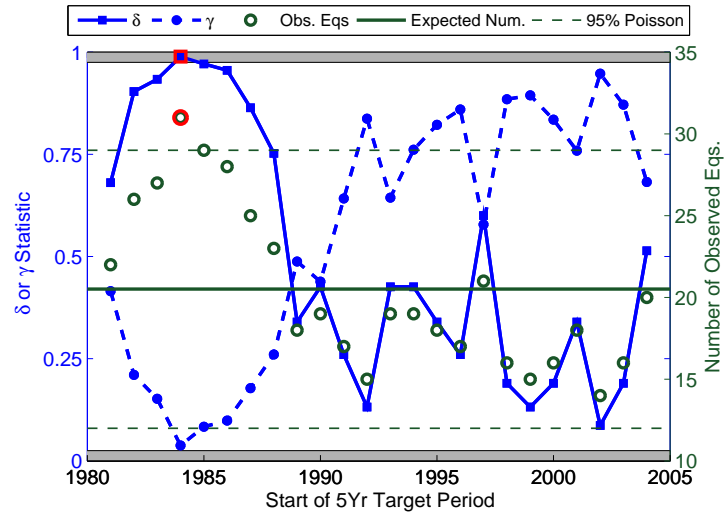}
\caption{\label{fig:RelmTestsMainshock} Retrospective consistency tests of the 5-year time-independent mainshock-only forecast on past, overlapping, declustered 5-year periods with starting year as abscissa. Left (blue) ordinate axis: The N-test statistic $\delta$ (blue squares), measuring the agreement between the number of expected and observed events, and the L-test statistic $\gamma$ (blue circles), measuring the agreement between the expected and observed likelihood score. Right (green) ordinate axis: Number of observed events (green open circles), the expected number of events (green solid line), along with $95\%$ Poisson confidence bounds (green dashed lines). Red outlines of the symbols mark periods during which the forecast is inconsistent with the observations. Grey bars denote rejection regions. 
}
\end{figure}

We also performed retrospective tests on our mainshock forecast (see Figure \ref{fig:RelmTestsMainshock}). In contrast to the all-earthquake forecast, this mainshock-only forecast passes the N-test in all (overlapping) target periods except the 1984-1989 period, during which the forecast underpredicted. The distribution of the number of mainshocks in five-year periods is clearly much more Poissonian than the number of all events. As a result, the Poisson forecast fares much better in this forecast category than in the all-earthquake group. Indeed, one would expect one rejection in 20 (non-overlapping) target periods at $5\%$ confidence. The forecast passes the L-test in every target period. This should not be surprising, given that the model is based on the spatial locations of the target catalogs (in this retrospective test). We did not apply the modified tests of section \ref{sec:NBD}.

\clearpage

\subsection{Supplement 4: Simplex Method for ETAS Parameter Estimation}

In this supplement, we discuss in more detail the performance of the ETAS model parameter estimation. The model's growing importance means that we need to thoroughly understand the likelihood estimation and associated uncertainties. The performance of the optimization is rarely discussed in the articles involving the ETAS model, although the parameter estimation is not simple. 

\begin{figure}[htbp]
\centering
\includegraphics[draft=\IsDraft,width=\mediumwidth,keepaspectratio=true,clip]{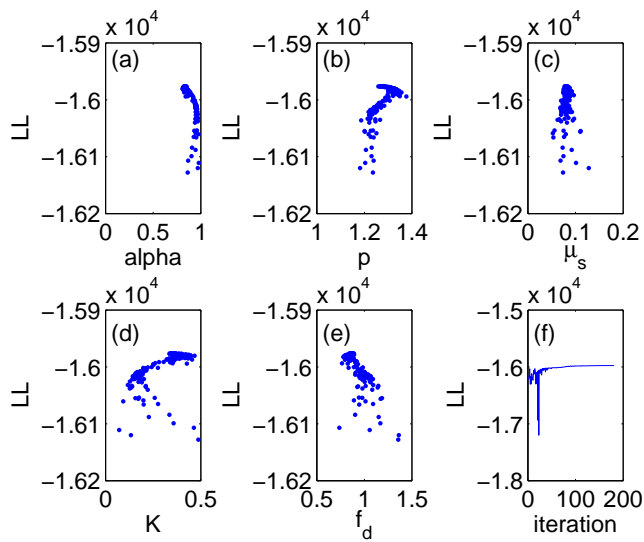}
\caption{\label{fig:LLprofile} Illustration of the simplex optimization method using model 1 in Table \ref{tab:optETASm4} with a Gaussian kernel: Value of the log likelihood as a function of each model parameter (a) - (e) and as a function of the number of iterations (f). Log likelihood profiles of the other models in Table  \ref{tab:optETASm4} are more Gaussian and less correlated. 
}
\end{figure}

\begin{figure}[ht]
\centering
\includegraphics[draft=\IsDraft,width=\halfwidth,keepaspectratio=true,clip]{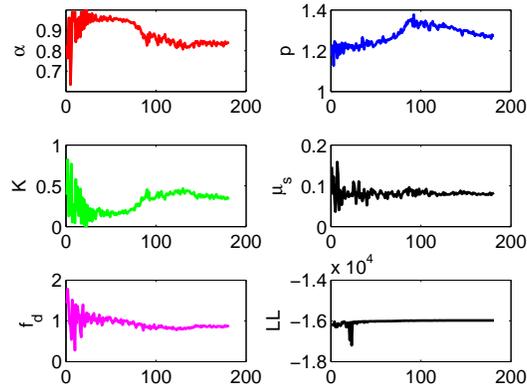}
\caption{\label{fig:ParmEvol} Evolution of the parameter estimates with the number of iterations for model 1 in Table  \ref{tab:optETASm4}. 
}
\end{figure}

\begin{figure}[ht]
\centering
\includegraphics[draft=\IsDraft,width=\halfwidth,keepaspectratio=true,clip]{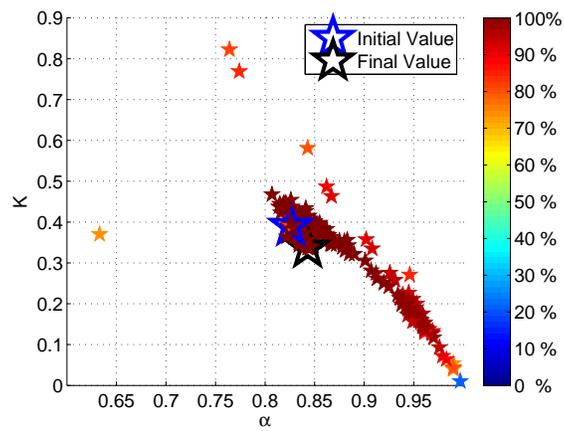}
\caption{\label{fig:AvK} Log-likelihood values (as a percentage of the final value) as a function of the ETAS model parameters $(\alpha, K)$ for all points visited by the simplex method during the optimization of model 1. The large blue star marks the initial values; the large black star the final estimates. 
}
\end{figure}

Figures \ref{fig:LLprofile} shows the log-likelihood scores as a function of parameter values visited during the simplex optimization. While the actual 5-dimensional likelihood function is difficult to visualize when projected onto each dimension individually, the correlations between the parameters are evident. The convergence of the parameters as a function of the iteration is shown in Figure \ref{fig:ParmEvol}. In Figure \ref{fig:AvK}, we projected the optimization steps of the simplex method onto the 2-dimensional log-likelihood surface as a function of $\alpha$ and $K$. 


Despite this evidence for a non-trivial likelihood surface near its maximum, the parameters of most models converged after about 100 iterations for all models for target events $m \ge 3.95$. This is in contrast to optimization for target events $m \ge 2$, which sometimes failed to converge after 200 iterations. For the latter target set, we also encountered cases where different random number generator seeds or parameter starting values led to significantly different parameter estimates (not shown). 

In Figures \ref{fig:LLprofile}, \ref{fig:ParmEvol} and \ref{fig:AvK}, we show the worst case scenario (models 1 and 9gs in Table \ref{tab:optETASm4}) of the optimization procedures to illustrate the difficulty of the non-linear, non-Gaussian estimation problem of the ETAS model parameters. Most of the other models could be estimated with a more well-behaved likelihood function. We are relatively confident that the simplex method performed adequately for the purposes of the CSEP target set of $m \ge 3.95$. Most likelihood profiles were better resolved around its maximum than in Figure \ref{fig:LLprofile}. In the future, we'd like to use a method that allows us to quantify uncertainties in the parameter values.

\clearpage
\subsection{Supplement 5: ETAS Parameter Optimization for Target Events $m \ge 2$}

To compare our parameter estimates to those obtained by \citet{Helmstetter-et-al2006}, we used the same target magnitude threshold $m_{min}=2$. Table \ref{tab:m2} presents the estimates, which are similar to those found by  \citet{Helmstetter-et-al2006}. However, the probability gains have universally decreased from the previous values above 10 to values closer to 5. There may be several reasons for the smaller gain. Firstly, \citet{Helmstetter-et-al2006} studied only southern California, which includes strongly clustered 1992 Landers to 1999 Hector Mine sequences. Expanding the region to all of California dilutes those gains, as more independent earthquakes are included. Of course, strong triggering sequences are also observed outside of southern California, but the statistics seem to be dominated by the strong activity in the Mojave Desert during the 90s. Secondly, our grid size is $0.1^\circ$ while \citet{Helmstetter-et-al2006} used a finer grid of $0.05^\circ$. A subtle, nonlinear trade-off probably exists between the realized gain of a time-dependent model and the effects of the spatial (and of course temporal) grid-size, the study region and the occurrence of large quakes and their triggered events. The sole objective measure of gain would be achieved on a global scale in continuous time. 


\begin{table}[htbp]
\caption{ {\bf Parameter estimates of the ETAS model for target threshold $m_{min}=2$:} \newline Input catalog: ANSS catalog 1/1/1981 to 1/4/2009 in the CSEP collection region (167,528 earthquakes $m\geq2$). Target catalog: ANSS catalog 1/1/1986 to 1/4/2009 in the CSEP testing region ($123,475$ events $m \geq 2$). Spatial background model 21 from Table \ref{tab:SmoOpt}. Reference model: time-independent forecast model 21 and the average number $\mu_{TI}=14.54$ of daily $m \ge 2$ earthquakes. $^{\ddagger}$: Using the corrective term $\rho^{\star}(m)$. $^{\S (m=M)}$: Using the early-aftershock-smoothing kernel for earthquakes above $m \geq M$. }
\begin{center}
  \begin{tabular}{@{} ccccccccccccc @{}}
    \hline
    Model & kr & $m_{min}$ & $\alpha$ & $p$ & $k$ & $\mu_s$ & $f_d$ & $L_{ETAS}$ &$L_{TI}$  & $N_{pred}$ & $G$ \\ 
    \hline
    $1$ 			& gs & 2.0 & 0.42 & 1.19 & 1.03 & 4.460 & 1.50 & -766,170 &  -975,835 &  123,532.2 & 5.46 \\  
    $2$ 			& pl & 2.0 & 0.38 & 1.21 & 1.21 & 4.046 & 0.002 & -765,130 &  -975,835 &  124,134.7 & 5.51 \\  
    $3^{\ddagger}$	& gs & 2.0 & 0.32 & 1.25 & 0.93 & 5.280 & 1.76 & -768,115 &  -975,835 & 123,658.5 & 5.38  \\  
    $4^{\ddagger}$	& pl & 2.0 & 0.33 & 1.25 & 0.98 & 4.916 & 0.006 & -767,433 &  -975,835 &  122,653.9 & 5.41  \\  
    $5^{m=5}$	         & gs & 2.0 & 0.42 & 1.20 & 1.07 & 4.485 & 1.56 & -766,035 &  -975,835 &  123,017.0 & 5.47 \\  
    $6^{m=5}$	         & pl & 2.0 & 0.39 & 1.21 & 1.19 & 4.105 & 0.015 & -765,181 &  -975,835 & 124,007.8 &  5.51 \\  
    $7^{m=6}$	         & gs & 2.0 & 0.40 & 1.22 & 1.14 & 4.554 & 1.41 & -766,258 &  -975,835 &  124,040.2 & 5.46 \\  
    $8^{m=6}$	         & pl & 2.0 & 0.38 & 1.21 & 1.21 & 4.057 & 0.02 & -765,144 &  -975,835 &  123,748.6 & 5.51 \\  
    
     \hline
  \end{tabular}
\label{tab:m2}
\end{center}
\end{table}

\end{document}